\documentclass[iop,apj]{emulateapj}
\usepackage{amsmath,amssymb,amstext,float}

\usepackage[breaklinks,colorlinks,citecolor=blue,linkcolor=magenta]{hyperref}

\usepackage[all]{hypcap} 
\usepackage{aas_macros}
\usepackage{natbib}
\usepackage{color}

\shorttitle{Coronal Heating}
\shortauthors{Knizhnik et al.}
\newcommand{\beg}[1]{\begin{equation}\label{#1}}
\newcommand{\done}{\end{equation}}
\newcommand{\pd}[2]{\frac{\partial #1}{\partial #2}}

\newcommand{\vecB}{\textbf{B}}

\newcommand{\vecS}{\textbf{S}}
\newcommand{\vecE}{\textbf{E}}

\newcommand{\vecv}{\textbf{v}}

\newcommand{\curl}[1]{\nabla\times{#1}}
\newcommand{\divv}[1]{\nabla\cdot{#1}}

\numberwithin{equation}{section}

\begin{document}

\title{The Role of Magnetic Helicity in Coronal Heating}
\author{K.\ J.\ Knizhnik\altaffilmark{1}, S.\ K.\ Antiochos\altaffilmark{2}, J.\ A.\ Klimchuk\altaffilmark{2} and C.\ R.\ DeVore\altaffilmark{2}}
\altaffiltext{1}{Naval Research Laboratory, 4555 Overlook Avenue SW, Washington, DC 20375 USA}
\altaffiltext{2}{Heliophysics Science Division, NASA Goddard Space Flight Center, Greenbelt, Maryland}

\begin{abstract}

  One of the greatest challenges in solar physics is understanding the
  heating of the Sun's corona. Most
  theories for coronal heating postulate that free energy in the form
  of magnetic twist/stress is injected by the photosphere into the
  corona where the free energy is converted into heat either through
  reconnection or wave dissipation. The magnetic helicity associated
  with the twist/stress, however, is expected to be conserved and appear in the corona. 
  In previous work we showed
  that helicity associated with the small-scale twists undergoes
  an inverse cascade via stochastic reconnection in the corona, and
  ends up as the observed large-scale shear of filament channels. Our
  ``helicity condensation'' model accounts for both the formation of
  filament channels and the observed smooth, laminar structure of
  coronal loops. In this paper, we demonstrate, using helicity- and
  energy-conserving numerical simulations of a coronal system driven
  by photospheric motions, that the model also provides a natural
  mechanism for heating the corona. We show that the heat
  generated by the reconnection responsible for the helicity
  condensation process is sufficient to account for the observed
  coronal heating.  We study the role that helicity injection plays in
  determining coronal heating and find that, crucially, the heating
  rate is only weakly dependent on the net helicity preference of the
  photospheric driving. Our calculations demonstrate that
  motions with 100\% helicity preference are least efficient at heating the corona;
  those with 0\% preference are most efficient. We
  discuss the physical origins of this result and its implications for
  the observed corona.
\end{abstract}

\maketitle


\section{introduction}\label{sec:intro}
Understanding the nature of coronal heating is a longstanding
problem in solar physics dating back to the work of
\citet{Grotrian39}, who discovered that coronal temperatures are in
excess of $10^6 \; \mathrm{K}$, approximately two orders of magnitude
hotter than the underlying photosphere. Calculations by
\citet{Withbroe77} showed that the energy flux necessary to account
for the observed temperatures is of order $1\times10^7 \;
\mathrm{ergs\; cm^{-2} \; s^{-1}}$ in active regions and $3\times10^5
\; \mathrm{ergs\;cm^{-2}\;s^{-1}}$ in quiet Sun. Explaining the
mechanism for the heating has been, to some extent, the central unsolved
problem in solar physics \citep{Parker72, Klimchuk06, Klimchuk15}. \par

While it is generally agreed that the energy must ultimately come from
the photosphere and below, there are two main mechanisms that
are frequently used to explain the observed heating. \citet{Parker72}
argued that quasi-static photospheric motions tangle, twist, and braid
the magnetic field, imparting it with magnetic free energy, which is
then converted into heat by magnetic reconnection. Such reconnection
events, frequently called nanoflares \citep[e.g.][]{Parker72,
Parker83, Cargill94, Klimchuk06, Klimchuk15}, are thought to occur
between elemental strands, perhaps as small as $15 \; \mathrm{km}$ in
diameter \citep{Peter13}.  An alternative explanation for the heating
is that waves, such as the Alfv\'en \citep{Osterbrock61} or
magnetosonic \citep{Pekunlu01} modes, are generated by
photospheric processes and deposit their energy into the corona by
resonant absorption, phase mixing, or turbulent dissipation. \par

These models are frequently able to quantitatively reproduce the
observed heating \citep[e.g.,][see, however,
\citealt{Gonzalez15}]{Kumar06, Viall11, Bradshaw12, Hahn14}, but
to date there has been little consideration of their topological
implications. Incessant jostling of the magnetic-field footpoints in
the photosphere causes the coronal field to become tangled and
twisted, thereby injecting helicity. There is, however, little
observational evidence for significant helicity, at least on large
scales. The corona has a rather simple, laminar appearance in images
ranging from the visible, to EUV, to soft X-ray
\citep{Schrijver99}. With very few exceptions \citep{Cirtain13}, coronal
loops --- defined here as observationally distinct structures --- seem
to be well aligned and to not wrap around each other. Complexity must
be present on sub-resolution scales, however, otherwise it would not be
possible to dissipate sufficient magnetic energy to account for the
observed coronal heating.  Furthermore, complexity may be present 
within the highly important diffuse component of the corona that
exists between loops \citep{Viall11}, but the level of topological
complexity that one might expect to accumulate on large observable
scales due to the injected helicity does not seem to exist. \par

One might argue that the expected complexity is simply removed by
magnetic reconnection. However, resolving the issue in this way is not
straightforward. In the high Lundquist-number regime of the solar
corona, magnetic helicity is conserved under reconnection
\citep{Woltjer58,Taylor74,Taylor86,Berger84b}. Reconnection rearranges
the helicity but does not destroy it. Some process must remove the
helicity from where it is locally injected, in order to avoid buildup
of tangling and braiding. We have recently described such a process. 
As a result of ``helicity condensation'' \citep{Antiochos13}, magnetic
twist naturally migrates via magnetic reconnection to polarity inversion
lines (PILs), where it manifests as sheared filament channels \citep{Zhao15}.
Using helicity-conserving MHD simulations, we showed that helicity 
condensation produces strong concentrations of shear at PILs, and 
leaves behind a generally smooth corona with well-aligned coronal 
loops \citep[][hereafter KAD15]{Knizhnik15}. Subsequently, we
demonstrated that this process occurs even if a fraction of the
helicity injected into the corona has a sign opposite to that of the
dominant helicity injection \citep[][hereafter KAD17]{Knizhnik17a}. 
Varying the fraction of positive/negative helicity affects both the
time scale of filament channel formation and the amount of structure
in the smooth-loop corona. In particular, if there
is a 3:1 ratio of positive/negative helicity, as appears to be the
case in the corona \citep{Pevtsov03}, filament channels will form in
about a day. On the other hand, a 1:1 ratio of positive/negative
helicity injection will preclude the formation of filament channels
altogether and, counter-intuitively, result in a more complex corona
than is generally observed. The helicity condensation process,
therefore, naturally explains why strong shear/twist is not observed
in the closed corona except in filament channels. \par

Although these results have shown that magnetic reconnection
transports magnetic helicity to PILs by removing it from the coronal
loops, it is not clear how much energy is converted to heat
in this process. Filament channels are highly sheared
structures that contain a large amount of free energy, which must
have come, at least in part, from the energy injected at the photosphere. 
\citet{Welsch15} measured the Poynting flux injected from the
photospheric level into the coronal field, and found values well in
line with the prediction of \citet{Withbroe77}. However, not all of
the energy injected is converted
to volumetric heating, as a large amount of the energy remains in the
coronal field in the form of filament channels. Furthermore, it is not obvious
that the conversion of magnetic energy into heat is
efficient enough to account for the observed heating. Quantifying precisely
how much of the Poynting flux is converted into heat by magnetic
reconnection, therefore, is crucial to testing the helicity
condensation model and understanding the energy source of the
multi-million degree corona. \par

Photospheric flows inject both magnetic helicity and free energy into
the coronal field. In contrast to helicity injection, which
depends only on the normal component of the magnetic field at the
photosphere, energy injection depends on both the normal and
horizontal components. In the
absence of flux emergence, the normal component of the magnetic field
can be readily determined, because it simply convects with the ideal
photospheric flow. The horizontal component of the magnetic field, on
the other hand, is much harder to measure observationally. Knowing the velocity
distribution on the photosphere is not sufficient to determine the
evolution of the horizontal component of the field, because magnetic
reconnection in the corona removes stress from the field, which is
then able to relax along the entire length of its field lines. This effect changes the
horizontal components of the field at the photosphere, implying that
the Poynting flux is not determined solely by the driver velocity, but
by a complicated feedback between photospheric motions and the coronal
response. Thus, while helicity injection can be computed directly from
the dynamics observed or imposed at the photosphere, energy injection
requires computation of the detailed coronal dynamics. \par

The interplay between photospheric stressing, magnetic reconnection,
and the subsequent energy release and relaxation has been investigated
by many authors. \citet{Rappazzo13} studied energy injection into a
plane-parallel coronal geometry by photospheric twisting motions. They
analyzed the turbulent cascade that resulted from a large-scale
photospheric driver and found that field lines developed a complex
twisted topology and released energy by reconnecting at many
small-scale current sheets. In addition to the transfer of energy from
large scales to small, they found that an inverse energy cascade also
took place. They demonstrated that on long time scales most of the
free magnetic energy was stored at the largest possible
scales. \citet{WS11} used resistive MHD simulations to investigate
braiding of a plane-parallel corona by photospheric motions and found
that heating of the corona via magnetic reconnection is crucially
dependent on the nature of these motions. Furthermore, they argued
that injecting large amounts of magnetic helicity does not
increase the amount of energy available for conversion to heat,
compared to injecting less helicity, because the resulting minimum-energy
state determined by helicity conservation will itself have a
higher energy. In all cases, however, the relaxation that follows
magnetic reconnection results in a complex distribution of current
sheets, which can themselves dissipate and release further energy
\citep{Pontin11,Dahlburg16}. \par

In this paper, we show that the magnetic reconnection responsible for
``condensing'' magnetic helicity onto the PIL converts the majority of
the injected magnetic energy into plasma heating at a rate sufficient
to sustain the observed coronal temperatures. Furthermore, we find the
surprising result
that the coronal heating due to this process is only weakly
dependent on the helicity injection preference, even though for
large net helicity injection a significant amount of magnetic energy
remains stored in the large-scale shear observed in filament
channels. This work, combined with the results of KAD15 and KAD17,
demonstrates that the helicity condensation model provides a
self-consistent explanation for both the structure and heating of
the corona. Magnetic energy and helicity are injected at the
photosphere, but whereas the helicity condenses to form
concentrated filament channels at PILs, magnetic energy is
everywhere converted into heat by magnetic reconnection. Thus, 
understanding the transport of magnetic helicity throughout 
the corona is crucial to understanding solar coronal heating. \par

The paper is organized as follows. In \S\ref{sec:model} we review the
setup of our numerical simulations. In \S\ref{sec:results} we discuss
the results, describing the partitioning of
injected energy into magnetic and internal energy and how the
resolution of the simulations affects our findings. In \S
\ref{sec:corona} we relate the numerical values from our simulations
to coronal measurements. Finally, in \S\ref{sec:discussion} we
present our conclusions and their implications for future
research. \par


\section{Numerical Model}\label{sec:model}
We solve the equations of magnetohydrodynamics (MHD) using the
Adaptively Refined Magnetohydrodynamics Solver
\citep[ARMS;][]{DeVore08} in three Cartesian dimensions. The equations
have the form
\beg{cont}
\pd{\rho}{t}+\divv{\rho\vecv}=0,
\done
\beg{momentum}
\pd{\rho\vecv}{t} + \divv{\left( \rho\vecv\vecv \right)} = - \nabla P + \frac{1}{4\pi} \left( \curl{\vecB} \right) \times \vecB,
\done
\beg{energy}
\pd{E}{t}+\divv{\left\{\left(E+P+\frac{B^2}{8\pi}\right)\vecv-\frac{\vecB(\vecv\cdot\vecB)}{4\pi}\right\}}=0,
\done
\beg{induction}
\pd{\vecB}{t} = \curl{\left(\vecv\times\vecB\right)},
\done
where
\beg{totenergydensity}
 E=U+K+M
\done
is the total energy density, the sum of the internal energy density
\beg{internal}
U=\frac{P}{\gamma-1},
\done
kinetic energy density
\beg{kinetic}
K=\frac{\rho v^2}{2},
\done
and magnetic energy density
\beg{magneticenergydensity}
M=\frac{B^2}{8\pi}.
\done
In these equations, $\rho$ is mass density, $T$ is temperature, $P$ is
thermal pressure, $\gamma = 5/3$ is the ratio of specific heats,
$\vecv$ is velocity, $\vecB$ is magnetic field, and $t$ is time. We
close the equations via the ideal gas law,
\beg{ideal}
P = \rho RT,
\done
where $R$ is the gas constant. \par

ARMS uses finite-volume representations of the variables to solve the
system of equations. Its Flux Corrected Transport algorithms
\citep{DeVore91} provide minimal, though finite, numerical
dissipation, which allows magnetic reconnection to occur. As a result,
ARMS conserves the magnetic helicity in the system to an excellent
approximation, even when reconnection occurs throughout a substantial
fraction of the total volume \citep[][]{Knizhnik15,Knizhnik17a,Knizhnik17b,Knizhnik18a}. \par

The homogeneity of the system of MHD equations
(\ref{cont})-(\ref{induction}) allows us to scale out characteristic
values of mass, length, and time (or equivalent combinations of these
fundamental units) and solve the equations in non-dimensional form. We
do this by extracting values of length $L_s$, mass density $\rho_s$,
and magnetic-field strength $B_s$ that can be specified later to
convert our non-dimensional numerical values into solar values
appropriate to different portions of the atmosphere (active region,
quiet Sun). \par

\begin{figure}[!p]
\centering\includegraphics[scale=0.5, trim=0.0cm 0.0cm 0.0cm 0.0cm]{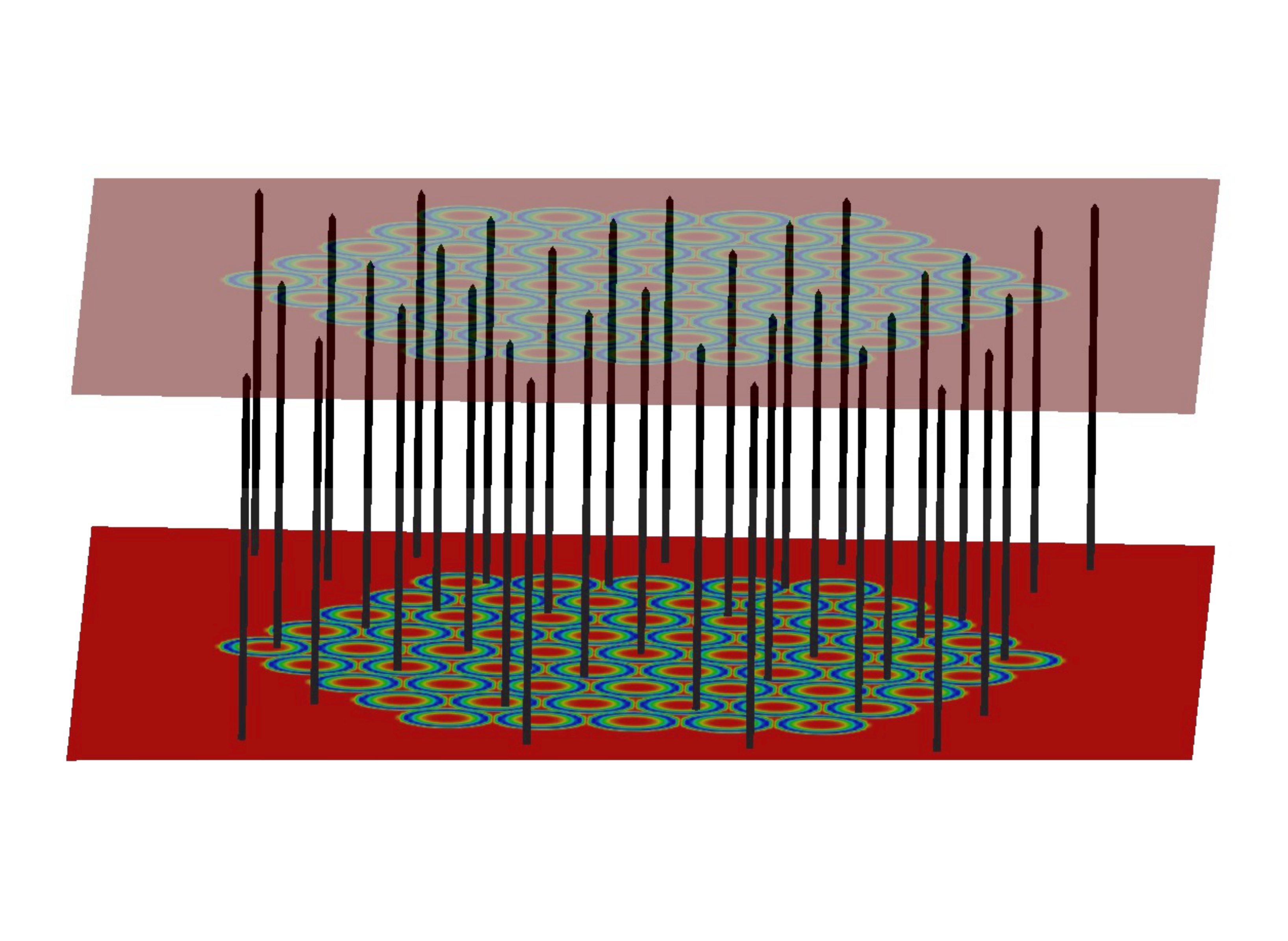}
\centering\includegraphics[scale=0.8, trim=0.0cm 0.0cm 0.0cm 0.0cm]{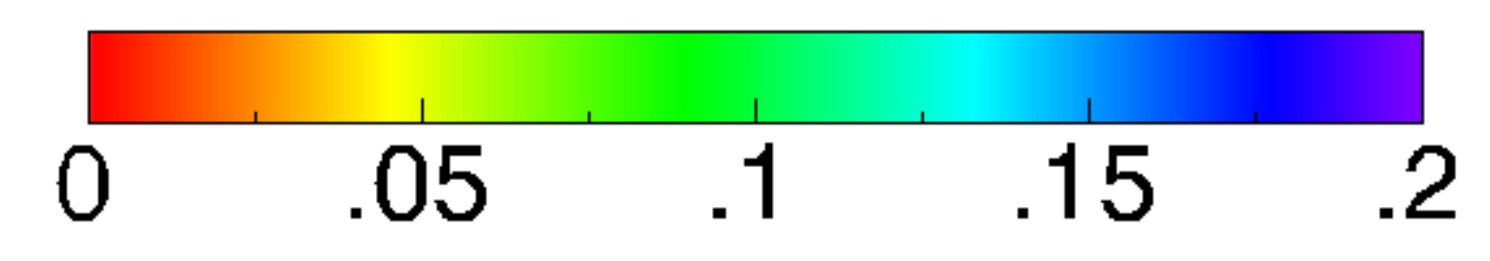}
\caption{Setup of the numerical simulations. Black lines represent the initial vertical magnetic field; color shading on the top and bottom plates represents velocity magnitude. To emulate the random photospheric convection in our numerical experiments, the local sense of rotation (clockwise/counter-clockwise) of individual convective cells shown in this figure is set randomly for helicity preferences $p \ne 100\%$, and the global hexagonal pattern of the cells collectively is rotated randomly about its center for all helicity preferences. }
\label{fig:init}
\end{figure}

Following Parker's (1972) original {\it ansatz}, we set up a model
coronal field that is initially straight and uniform between two
plates, as shown in Figure \ref{fig:init}. In this model, straight
flux tubes represent coronal loops whose apex is located in the center
of the domain, and the top and bottom boundaries represent the
photosphere. At all six boundaries, we apply zero-gradient conditions
to $\rho$, $E$, and $\vecB$. The four side walls are open, with
zero-gradient conditions applied to $\vecv$, but due to the form of
the driver discussed below, we find that the velocities at these side
walls are negligible throughout all the simulations. The top and
bottom boundaries are closed, where we set the normal flow component
$v_x=0$ and specify the tangential components $v_y, v_z$ as described
below. These imposed boundary flows mimic driving at the
plasma-dominated photosphere where the magnetic field is
line-tied. \par

Our domain is $[0,L_x] \times [-L_y,L_y] \times [-L_z,L_z]$,
where $x$ is normal to the photosphere (the vertical direction). We
set $L_x=1$ and $L_y=L_z=1.5$. The domain is spanned initially by a
grid of $4 \times 12 \times 12$ blocks, where each block consists of
$8 \times 8 \times 8$ finite-volume cells. The grid was refined up to
three additional levels above this base resolution, for four total
refinement levels $l$ = 1, 2, 3, and 4. The main results presented in
the next section were obtained on the $l$ = 3 grid, whose effective
uniform size was $128 \times 384 \times 384$. \par

We set the initial, uniform values in our non-dimensional simulations
to $\rho_0=1$, $T_0=1$, $P_0=0.1$, and $B_0=\sqrt{4\pi}$. These
choices set the sound speed, $c_{S0}=\sqrt{\gamma P_0/\rho_0}=0.4$,
Alfv\'en speed, $c_{A0}=B_0 / \sqrt{4\pi\rho_0} = 1$, and plasma beta
(ratio of thermal to magnetic pressure), $\beta_0=8\pi
P_0/B_0^2=0.2$. $\beta \ll 1$ corresponds to a magnetically dominated
plasma, and this condition generally holds in the corona. The results
presented below are given in simulation time, which is normalized
to the (unit) time required for an Alfv\'en wave at ambient speed to
propagate between the top and bottom plates
($t_{A0}=L_x/c_{A0}=1$). \par

As with our earlier studies, we inject helicity in the form of
localized vortices over a finite region to approximate a finite polarity
region bounded by a PIL.  Such vortices are observed in the
photosphere at the boundaries between convection cells
\citep{Duvall00,Gizon03,Komm07,Bonet08,Seligman14,VD15}.  Furthermore,
turbulent photospheric convection produces random flows, which can
always be deconvolved into translational and rotational
components. For simplicity, we approximate these flows with a
rotational pattern, described below, that allows for easy adjustment of
the helicity injection rate while maintaining the random nature
of the flow. As in KAD17, we impose a photospheric hexagonal pattern
of $N=61$ cellular rotations (shown in Fig.~\ref{fig:init}) that is
randomly shifted by an arbitrary angle after each complete cycle of
twist. Each such cycle consists of a sinusoidal ramp-up and ramp-down,
with the angular velocity of each individual cell taking the form
\beg{Omega} \Omega(r,t) = -\Omega_0g(r)f(t) \done where \beg{gofr}
g(r) = \left(\frac{r}{a_0}\right)^4 - \left(\frac{r}{a_0}\right)^8
\done and \beg{foft} f(t) =
\frac{1}{2}\left[1-\cos\left(2\pi\frac{t}{\tau}\right)\right].  \done
Here, $r$ is the cylindrical coordinate on the plane boundary measured
from the center of the cell, limited to the interval $r \in [0,a_0]$ in
Eq.\ \ref{gofr}. {\color{black} This flow is incompressible and leaves 
the vertical magnetic field component $B_x$ undisturbed; the tangential 
field component {\bf B}$_h$, on the other hand, is free to adjust in 
response to the imposed surface flows and to the dynamics occurring 
in the coronal volume above. The zero-gradient conditions allow 
any change in {\bf B}$_h$ in the interior to propagate to and appear at the boundary, 
so that any accumulation of {\bf B}$_h$ at the boundaries and in the volume is fundamentally 
due to coronal reconnection and not a direct result of boundary motions.} We set the flow amplitude
$\Omega_0 = 7.5$, the flow radius $a_0=0.125$, and the flow period
$\tau=3.35$, which yield a maximum angle of rotation within the cell
$\phi_{max} = \pi$ and a maximum flow speed $v_h^{\rm max}=0.2$. The
twisting is applied on both the top and bottom plates, so the maximum
rotation angle of each flux tube is $2\pi$. Each simulation is run out
for $20$ twist cycles, which are followed by $5$ relaxation cycles
during which no further twisting motions are applied. \par

In addition to randomly rotating the hexagonal pattern from cycle to
cycle, two of our simulations further randomize the energy and
helicity injection into the corona by randomizing the sense of
rotation (clockwise or counter-clockwise) of the individual cells
after each cycle. As described in KAD17, our three simulations
presented below correspond to net preferences $p$ of $100\%$, $50\%$,
and $0\%$ for clockwise (positive-helicity) rotations. In the first
case, all rotations are clockwise; in the third case, half are
clockwise and half counter-clockwise; and in the second, the ratio of
clockwise to counter-clockwise rotations is 3:1. A key point is that
even though our imposed motions consist of simple rotations, the
random rotation of the flow pattern implies that the effective driving
is fully chaotic, like actual solar convective flows. Any one
point at the photosphere moves along an unpredictable chaotic path. 
Consequently, our imposed driving captures the essential features
of the convective turbulence
while allowing for simple adjustment of the helicity injection
rate. \par

The key difference between the present study and our previous work
(KAD15, KAD17) is that here we solve Eq.\ \ref{energy} to conserve
total energy during the evolution. In our previous simulations, we
solved the adiabatic temperature equation, and all changes in the
internal energy density were due to compression or expansion. These
two descriptions are equivalent if the evolution is purely
ideal. However, the numerical diffusion terms in the momentum and
induction equations provide finite viscous and resistive damping,
respectively, that produces localized heating. With the adiabatic
approximation these heating terms are simply lost. By solving the
conservative form of the total energy equation, on the other hand, we
capture the full heating. Our energy equation is still an
approximation for the actual corona: no thermal conduction or
radiation is included in our simulations. Consequently, all
temperature increases due to the viscous and resistive heating simply
accumulate monotonically. Nevertheless, by capturing rigorously the
conversion of kinetic and magnetic energy into plasma energy during
reconnection, we are able to assess quantitatively the amount of
coronal heating that occurs in our simulations and how that varies for
different helicity injection preferences and other parameters. \par


\section{Results}\label{sec:results}
We present several diagnostics that quantify the heating of the
corona for each of our simulations with varying helicity injection
preferences, beginning with the final temperature
distributions. Then we focus on the energetics addressing, in turn,
the Poynting fluxes and the volumetric energies -- magnetic, internal,
and total. Finally, we show and discuss the effects of grid resolution
(i.e., the effective Lundquist number) on our results. \par


\subsection{Temperature}\label{sec:Temp}
\begin{figure*}[!p]
\centering\includegraphics[scale=0.4,trim=0.0cm 2.5cm 2.0cm 2.0cm, clip=true]{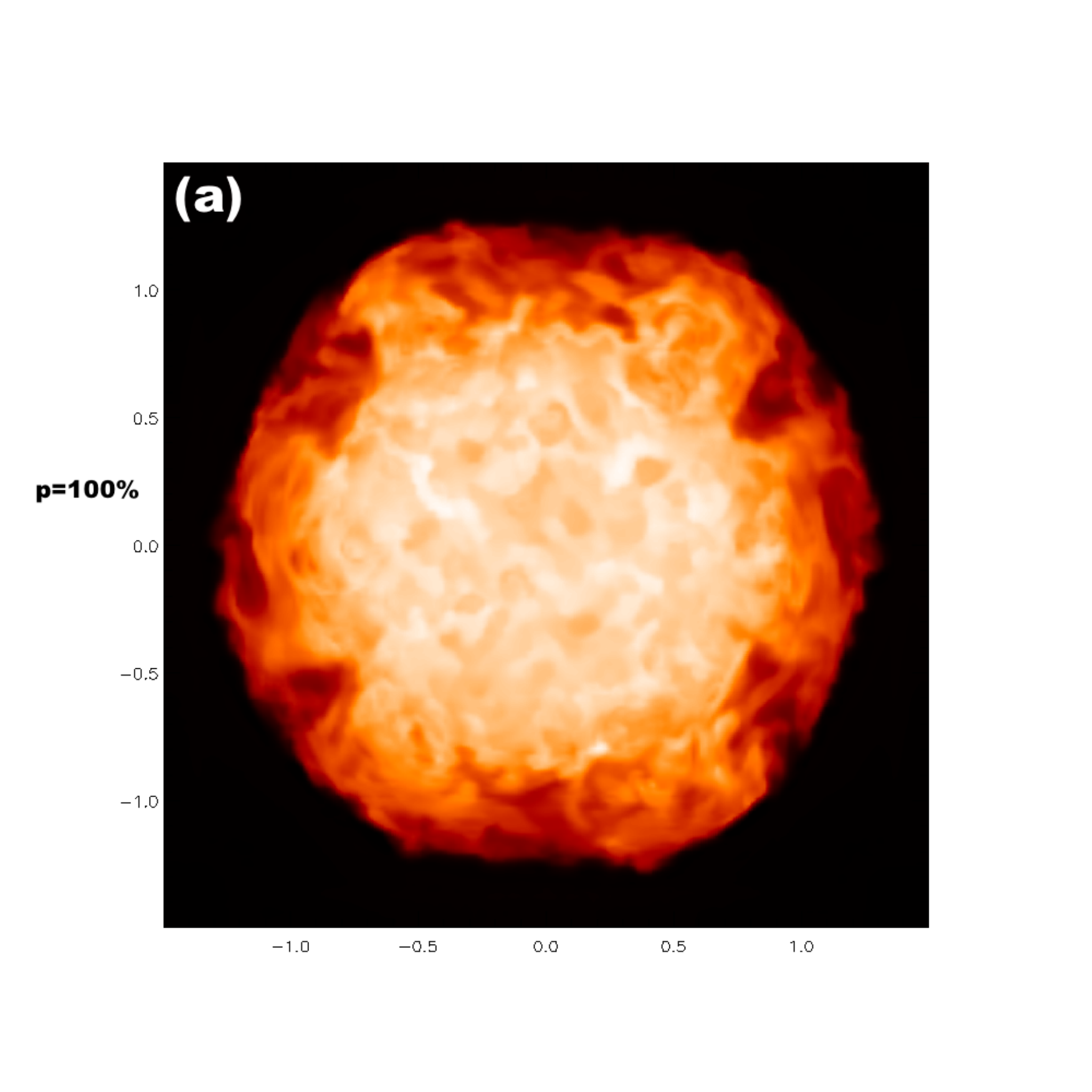}
\centering\includegraphics[scale=0.4,trim=2.5cm 2.5cm 0.0cm 2.0cm, clip=true]{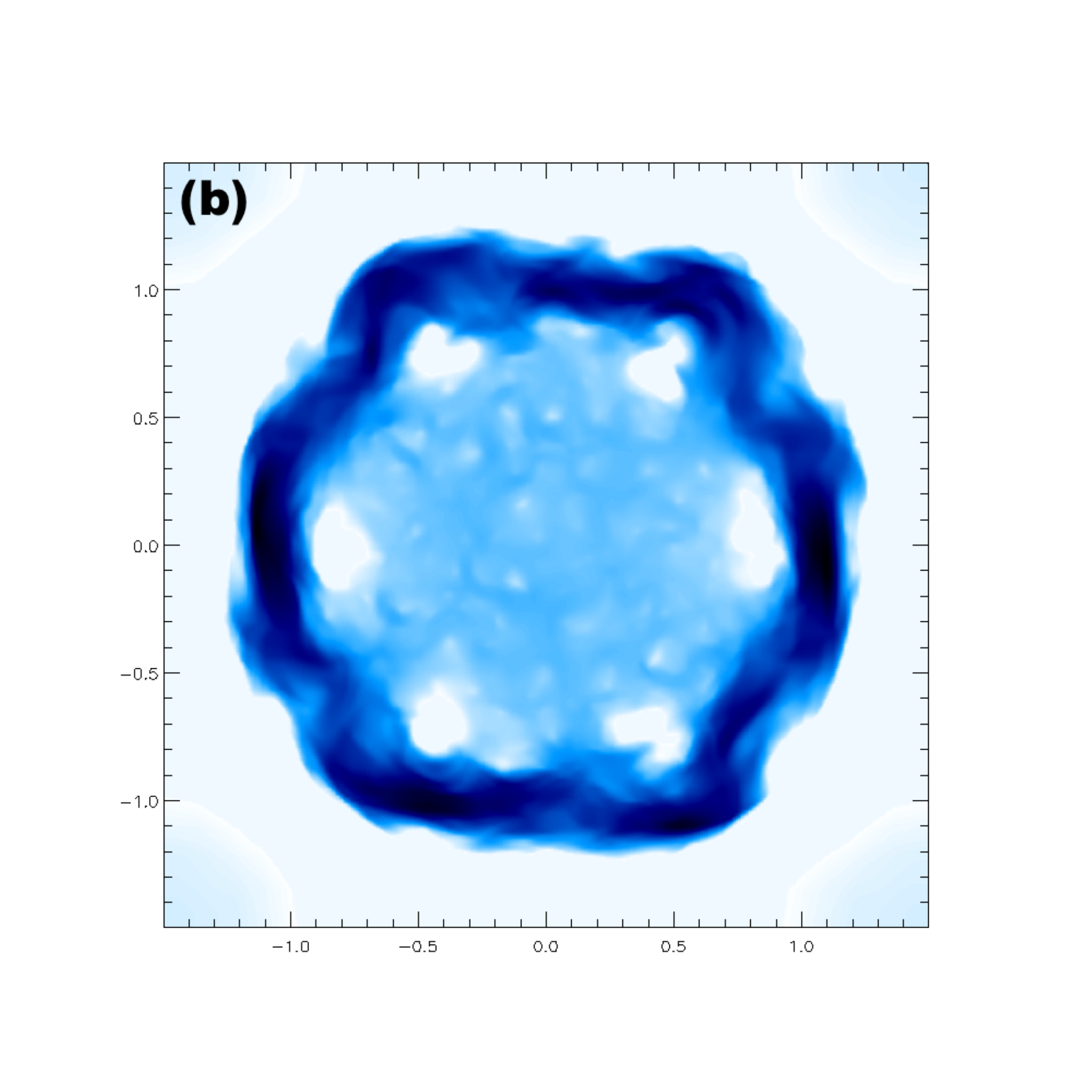}
\centering\includegraphics[scale=0.4,trim=0.0cm 2.5cm 2.0cm 2.0cm, clip=true]{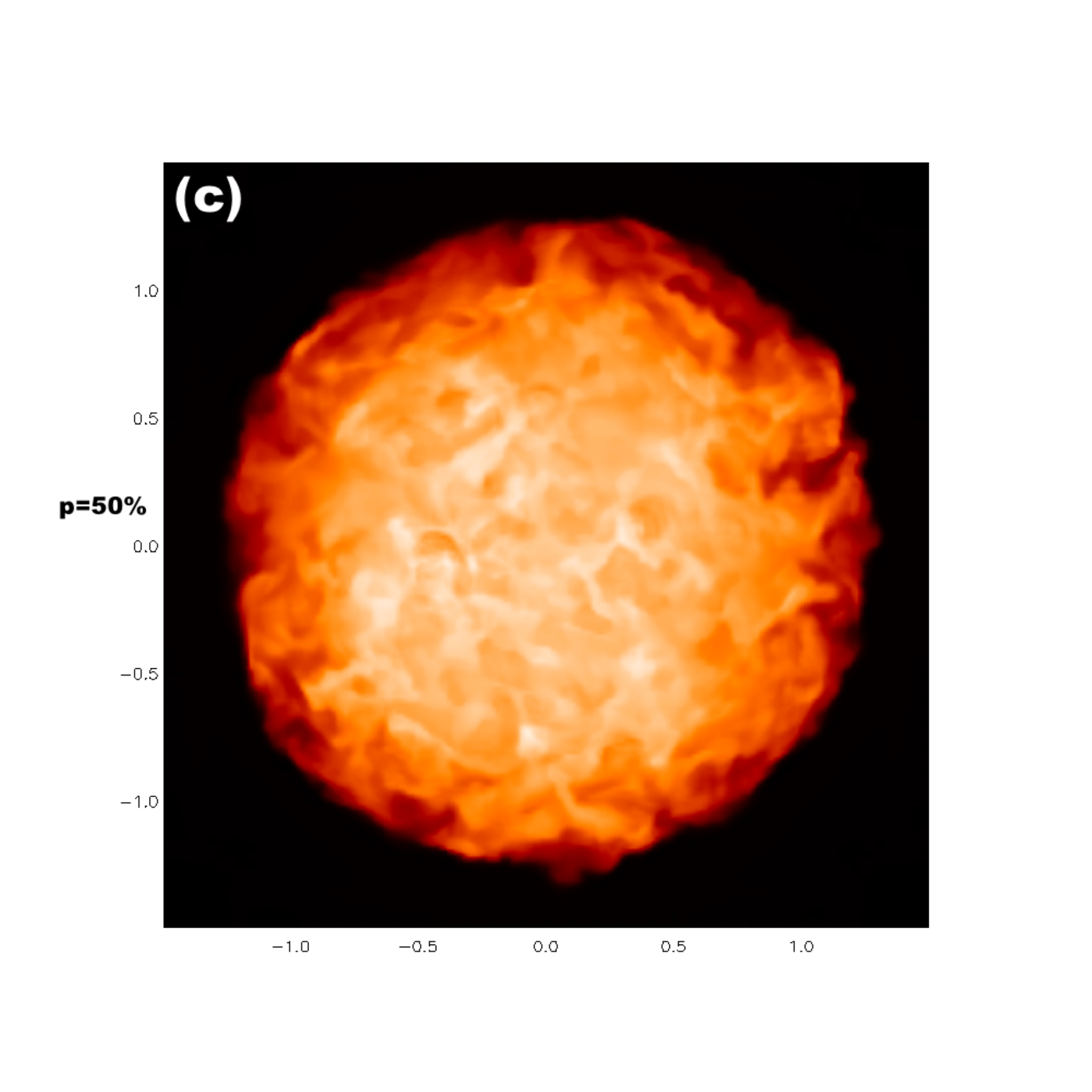}
\centering\includegraphics[scale=0.4,trim=2.5cm 2.5cm 0.0cm 2.0cm, clip=true]{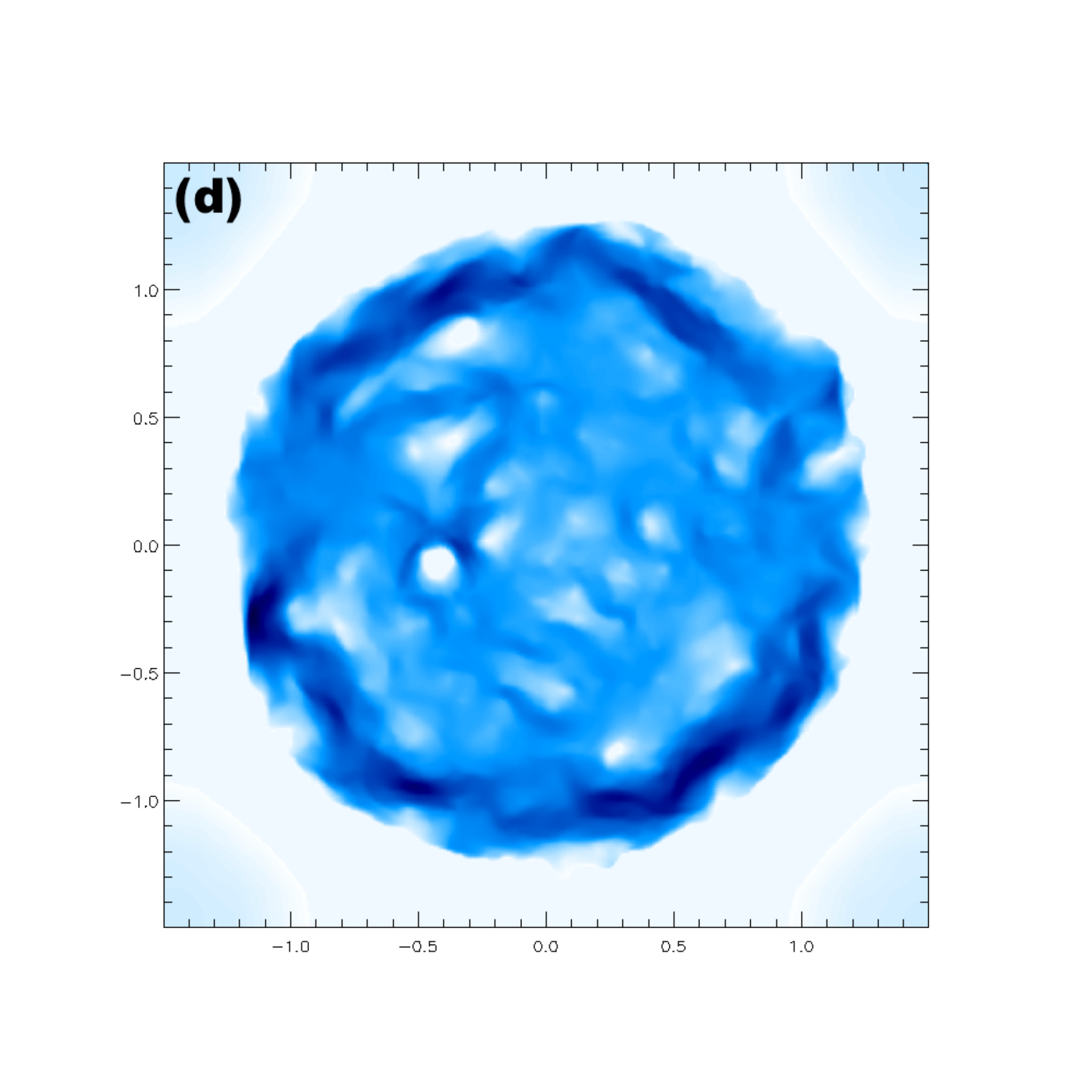}
\centering\includegraphics[scale=0.4,trim=0.0cm 2.5cm 2.0cm 2.0cm, clip=true]{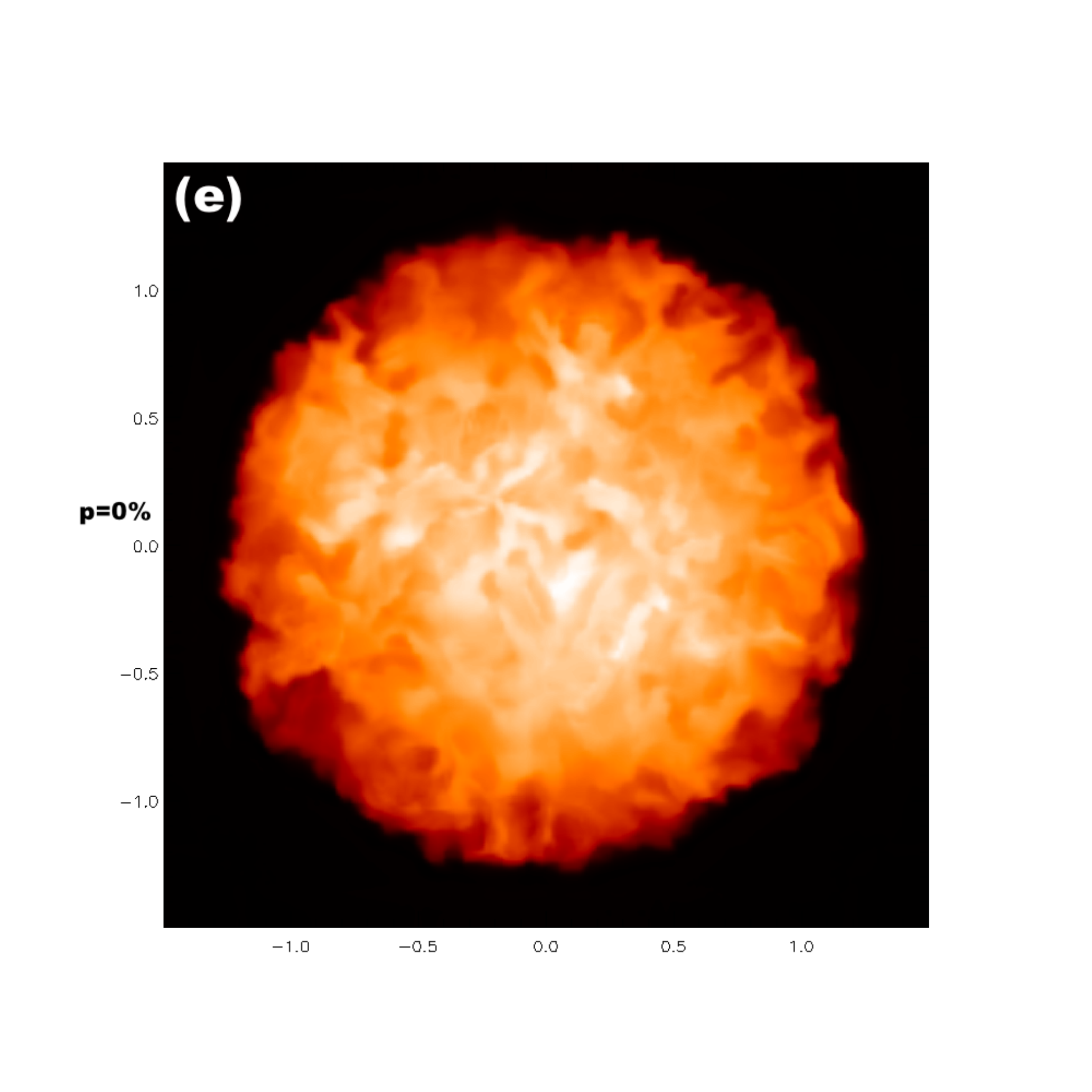}
\centering\includegraphics[scale=0.4,trim=2.5cm 2.5cm 0.0cm 2.0cm, clip=true]{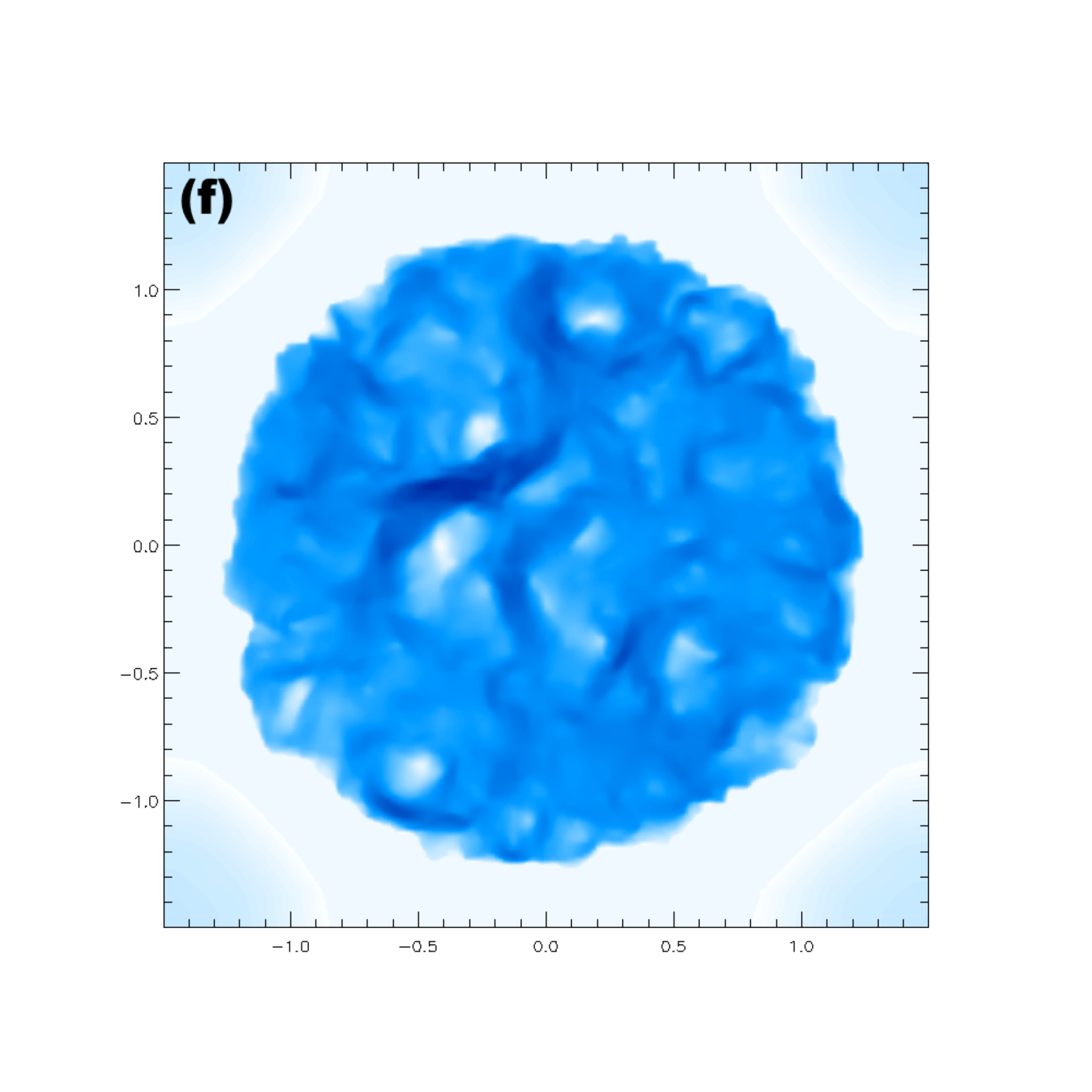}
\centering\includegraphics[scale=0.2,trim=5cm 0cm -30.0cm 0cm]{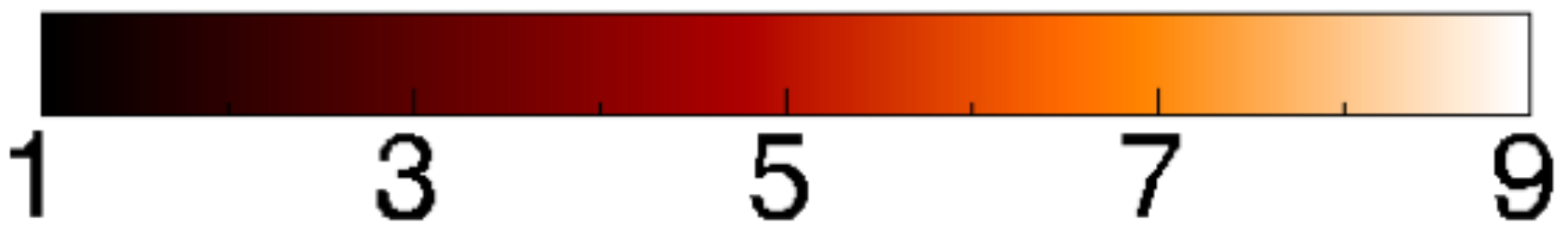}
\centering\includegraphics[scale=0.3,trim=10cm 0cm 4.0cm 0cm]{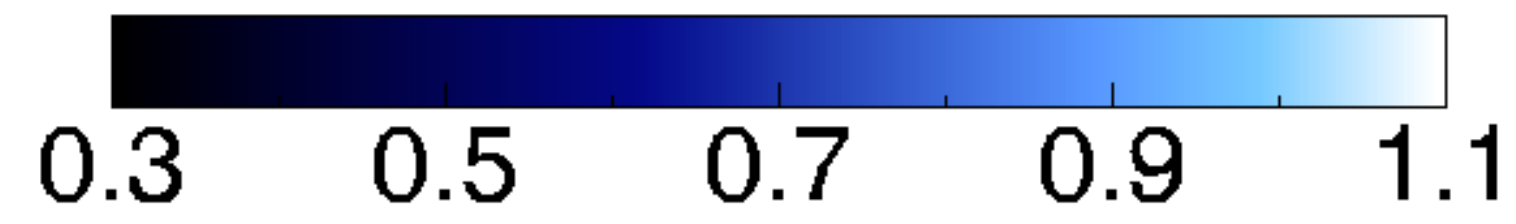}
\caption{Temperature (left) and $B_x/B_{0}$ (right) in the midplane at the end of the simulation for the $p=100\%$ (top), $p=50\%$ (middle), and $p=0\%$ (bottom) preferences. }
\label{fig:temperature}
\end{figure*}

As an initial assessment of the coronal heating implied by the
helicity condensation model, we plot the temperature in the
mid-plane of the domain at the end of each simulation in
Figure \ref{fig:temperature}. The contours show the ratio of the
final temperature to the initial temperature, $T_0 = 1$. Since we
apply closed boundary conditions at the top and bottom of the domain,
and the motions at the side walls remain very small, the total mass
is conserved throughout the evolution. Consequently, any
heating results primarily in an increase in temperature. The
temperature map, therefore, is a fairly accurate indicator of the
spatial distribution of the time-integrated heating over the
domain. There are some changes in plasma density, however, due to the
formation of the filament channel and the resulting expansion of the
flux in that region balanced by compression in the surrounding. To
illustrate those changes, we also show in Figure \ref{fig:temperature}
the vertical magnetic field at the midplane relative to its initial
value. Any increase (decrease) of this field from its initially uniform
strength $B_0=\sqrt{4\pi}$ indicates a local contraction (expansion)
there.\par

The first feature to note is that, in all cases, the temperature and the
vertical field exhibit a sharp jump between the driven and undriven
regions. This is to be expected given that the driving pattern of
Figure \ref{fig:init} has a nearly discontinuous boundary for any one
rotation cycle. The fact that we randomly rotate the whole pattern
creates a finite interface between the driven and undriven regions,
but its width is only of order a fraction of a rotation cell. 
Figure \ref{fig:temperature} shows that there is minimal spreading
of the injected twist field, whether due to reconnection or to numerical
diffusion. \par

The second important feature is that the temperature reaches its maximum
near the center of the driven region, and it is approximately the same
in all three cases. This result implies that the amount of magnetic
energy dissipated by reconnection is roughly constant, even though the
net helicity injection rate is very different in the three cases. We
discuss this key finding in more detail below. Furthermore, in all
cases the central region does not exhibit any pronounced large-scale
structure. Two caveats, however, need to be noted here when applying
our results to the actual corona. First, our model corona has
artificial uniformity in that the normal field at the photosphere is
constant and all the field lines have the same length. In the true
corona, both the photospheric field and the field-line length vary,
which naturally results in observed systematic temperature
variations. Second, there is no heat loss via radiation in our corona,
so that the plasma beta becomes artificially large. From
Figure \ref{fig:temperature}, we note that that the temperature at the
center has increased by almost an order of magnitude, which implies
that the beta there has risen significantly above unity. In the real
corona, the plasma is always low beta and is continuously
cooling and heating, which at any instant results in strong
cross-field temperature variations. \par

On the other hand, the lack of substantial structure in the
magnetic-field contours agrees well with the many observations that the
coronal field appears largely smooth and laminar \citep{Schrijver99}.  The
central portion of our driven region corresponds to the bulk of the observed
closed corona: coronal loops outside of filament channels. It is
evident from Figure \ref{fig:temperature} that the central region is even
smoother in the case with 100\% helicity injection than in that with
net zero helicity injection. This may seem counter-intuitive, because
one might expect to see evidence of the large helicity injection. In
fact, however, we have found (KAD17) that a 100\% helicity
injection preference consistently leads to a coronal magnetic field
outside filament channels that exhibits the least amount of
large-scale structure. This occurs because, in the 100\% case, the
helicity condensation process is most efficient at transporting the
injected magnetic twist to the PIL, where it builds up the sheared
filament channel. \par

Although the central parts of the driven regions are fairly similar
in the three cases, the boundaries clearly exhibit systematic
differences. {\color{black} This is to be expected, because these
boundaries separate the twisted and untwisted flux systems, and 
so act analogously to PILs:} filament channels can form there, but 
only if a nonzero net helicity is injected into the
corona \citep[e.g.,][]{Zhao15, Knizhnik15, Knizhnik17a, Knizhnik17b}.
First, we note from Figure \ref{fig:temperature} that throughout the
driven region, the vertical field $B_x$ is smaller than its initial value
$B_0 = 1$. This is due to significant horizontal field components,
$B_y$ and $B_z$, that have been generated in this region by the
photospheric motions. The extra magnetic pressure contributed by the
horizontal field causes the driven region to expand outward,
especially at the midplane, thereby lowering the average vertical field
strength. For the 0\% helicity injection case, the vertical field
is roughly uniform throughout the driven region, indicating that the
magnitude (but not the sign) of the horizontal field is also roughly
uniform there. The 100\% case, however, shows a much different
structure. The {\color{black} vertical} field exhibits a ring of large expansion at the
boundary -- the ``PIL'' -- due to the buildup of very strong
horizontal field, i.e., the sheared filament channel. In fact, 
the maximum magnitude of the horizontal field is more
than twice as large as the local vertical field magnitude. 
This ratio is in line with that in true filaments, where the {\color{black} horizontal} field is
measured to be up to an order of magnitude larger than the vertical
field \citep{Mackay10}. Due to the concentration of the shear field in the
filament channel, the interior region of the 100\% case exhibits less
expansion and weaker horizontal field than in the 0\% case, showing again
that the 100\% case is most efficient at eliminating the horizontal
(shear) flux by condensing it at the PIL. The ring of large field
expansion results in a ring of lower temperature at the boundary of
the 100\% case. (We emphasize here that this temperature decrease has no
relation to the physical process by which filament channels acquire cool
plasma.)  Note also that the 100\% case exhibits a weak remnant of the
hexagonal photospheric driving pattern (Fig.~\ref{fig:init}) in the
magnetic field and temperature. This weak effect is not present in 
the other cases due to the greater randomness of their photospheric
twisting motions. \par



\subsection{Poynting Flux}\label{sec:PF}
The imposed footpoint motions in our simulations inject energy
principally into the magnetic field. There is an attendant injection
of kinetic energy associated directly with the motions, but this
energy contribution is very small, of order 10$^{-4}$ times
that of the magnetic energy. The Poynting flux density of
electromagnetic energy (i.e., energy per unit area per unit
time) at any location within the domain is
\beg{PFeq}
\begin{split}
\vecS &= \frac{c}{4\pi} \left[ \vecE \times \vecB \right] \\
      &= - \frac{1}{4\pi} \left[ \left( \vecv \times \vecB \right) \times \vecB \right].
\end{split}
\done
The fluxes at the domain boundaries determine the energy added
to, or subtracted from, the total volume. The fluxes at the
side walls are negligible, because the outflows there are very small.
The fluxes at the top and bottom boundaries are due solely to the
imposed horizontal footpoint flows $\vecv_h$, because the normal
velocity component $v_n$ is held fixed at zero. Thus, the normal
component of the Poynting flux density at the top and bottom
boundaries is given by
\beg{Poynting} S_n(y,z,t) = -\frac{1}{4\pi}
B_n \left( \vecv_h \cdot \vecB_h \right),
\done
and it is the source of essentially all the energy added to the volume
during the simulation. Note also that this energy is all free energy
because the imposed motions are incompressible, so the uniform
normal flux at the boundary does not change throughout the simulation. 
Hence, the reference potential field is constant. \par

\begin{figure*}[!p]
\centering\includegraphics[scale=0.4,trim=-5.0cm 1.0cm 3.0cm 0.0cm, clip=true]{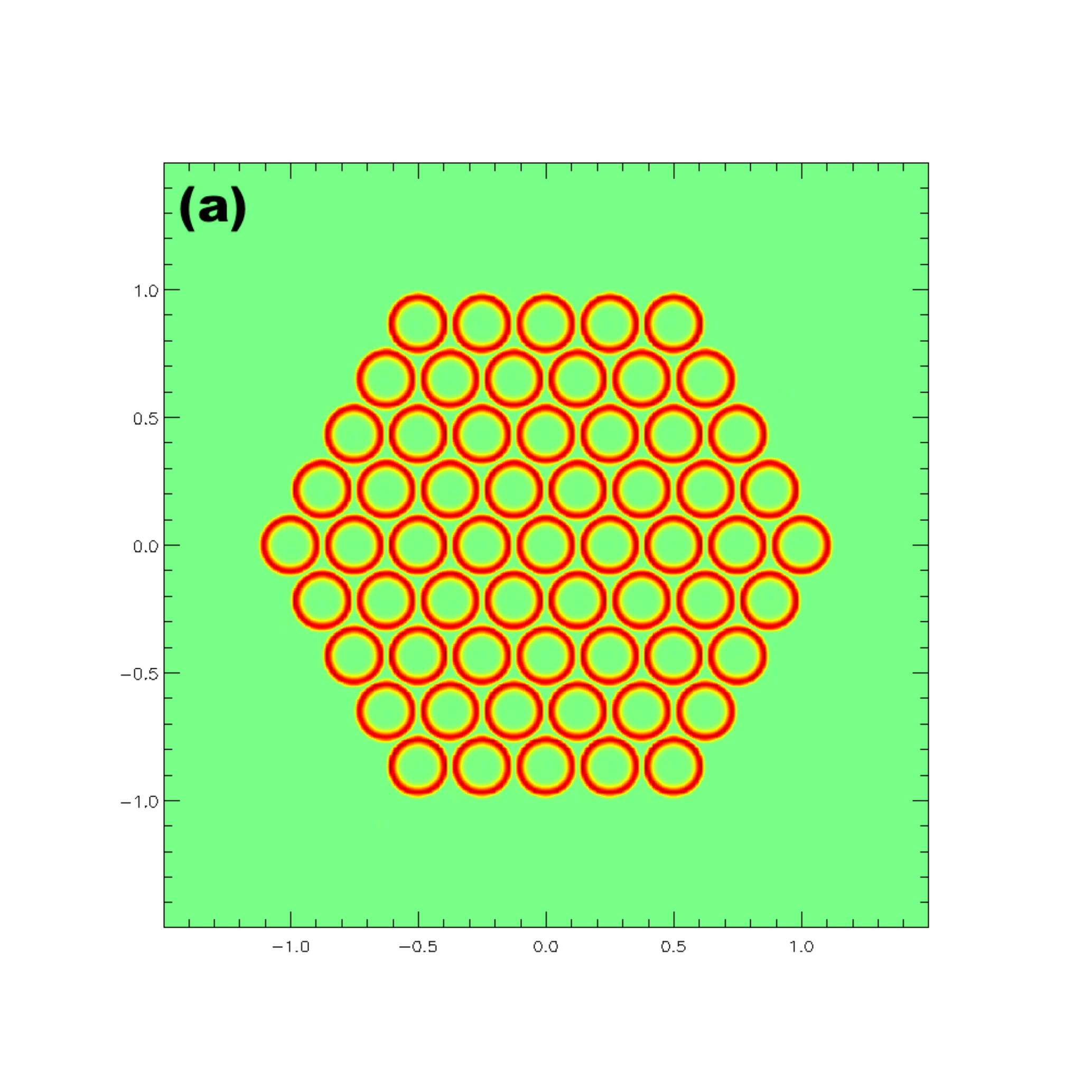}
\newline
\centering\includegraphics[scale=0.34,trim=-8.0cm 0.0cm 3.0cm 2.0cm]{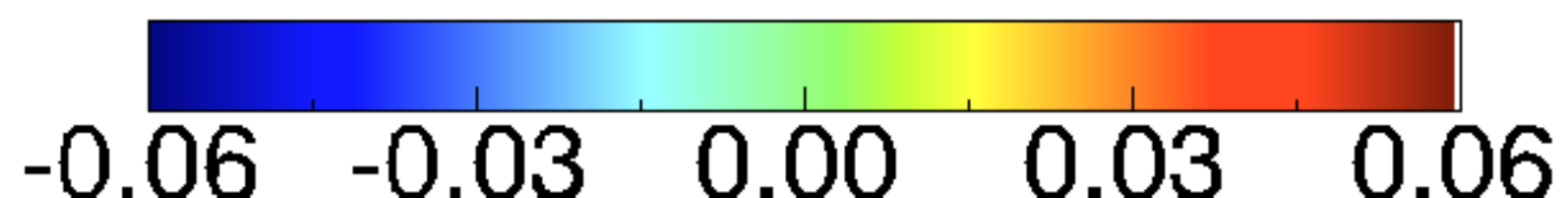}
\newline
\centering\includegraphics[scale=0.4,trim=2.0cm 2.0cm 2.0cm 2.0cm, clip=true]{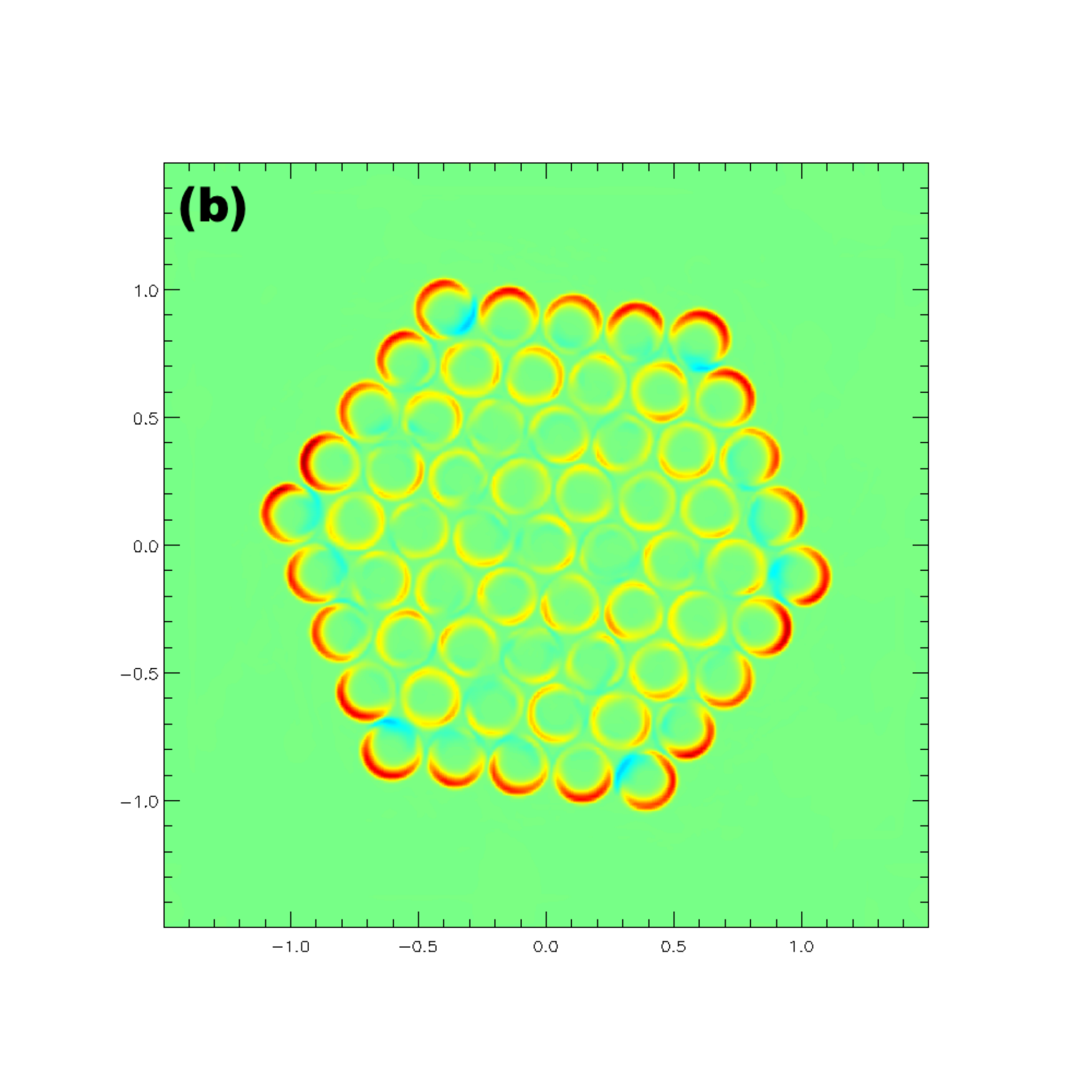}
\centering\includegraphics[scale=0.4,trim=2.0cm 2.0cm 2.0cm 2.0cm, clip=true]{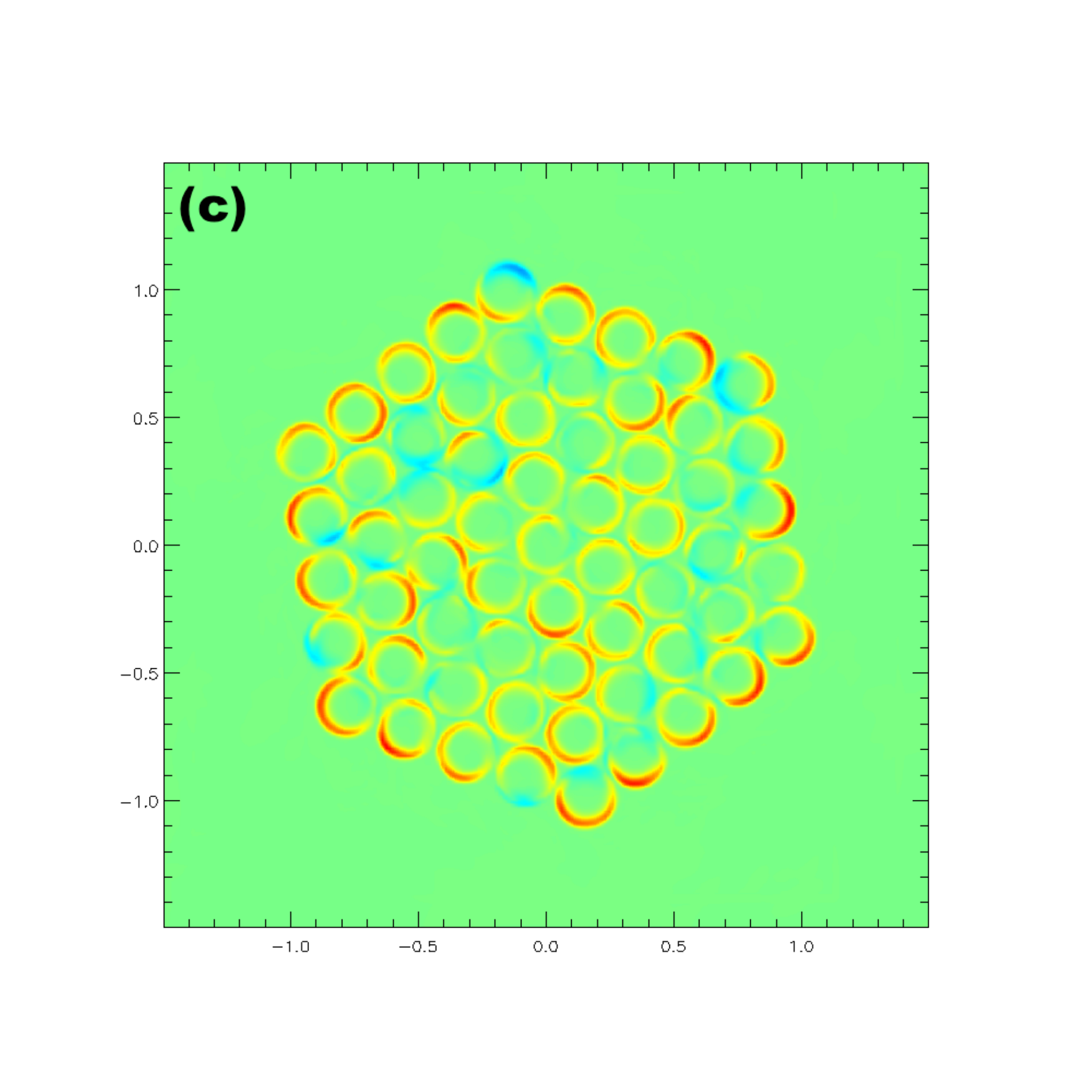}
\centering\includegraphics[scale=0.4,trim=-5.0cm 0.0cm 2.0cm 3.0cm, clip=true]{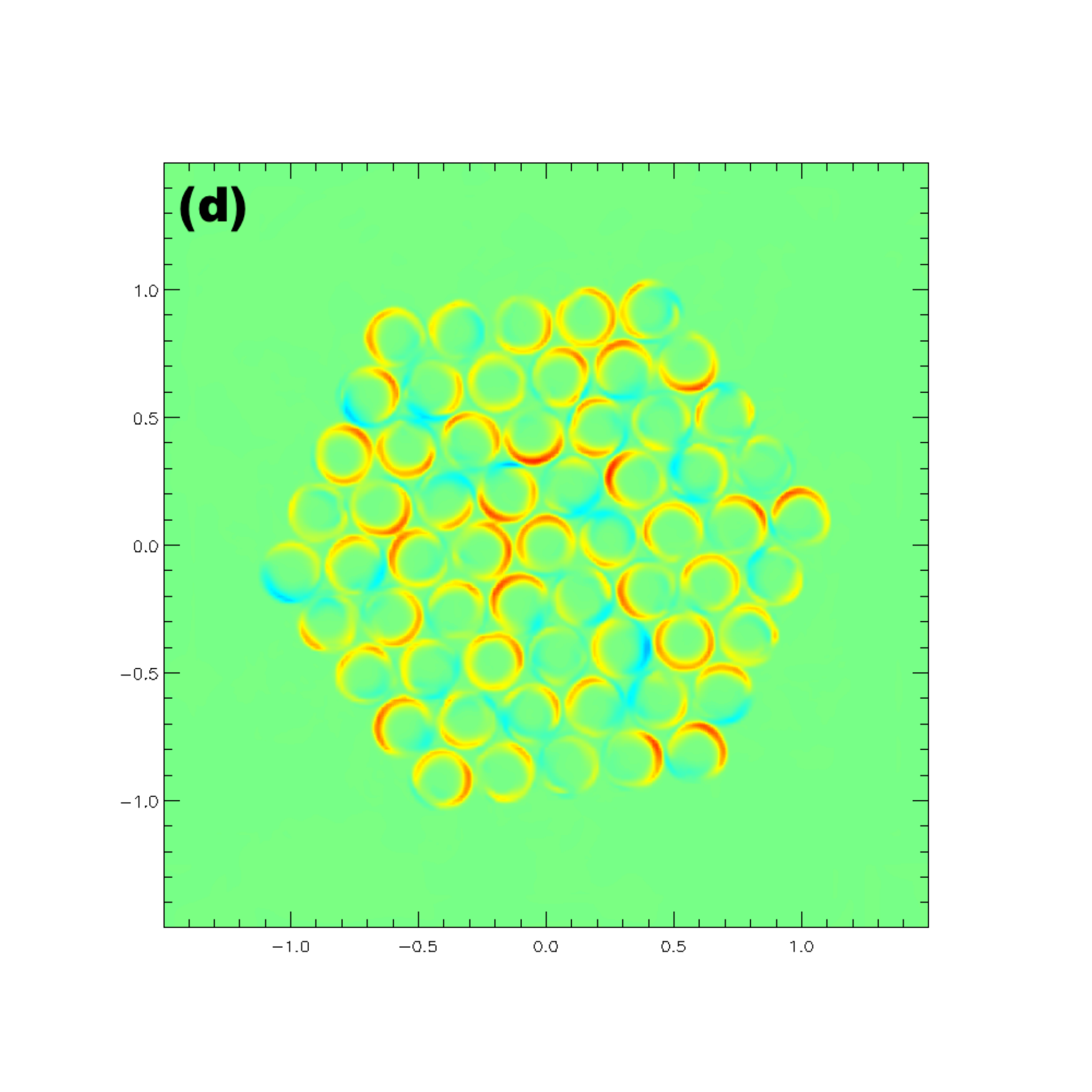}
\newline
\centering\includegraphics[scale=0.34,trim=-1.0cm 0.0cm 0.0cm 6.0cm]{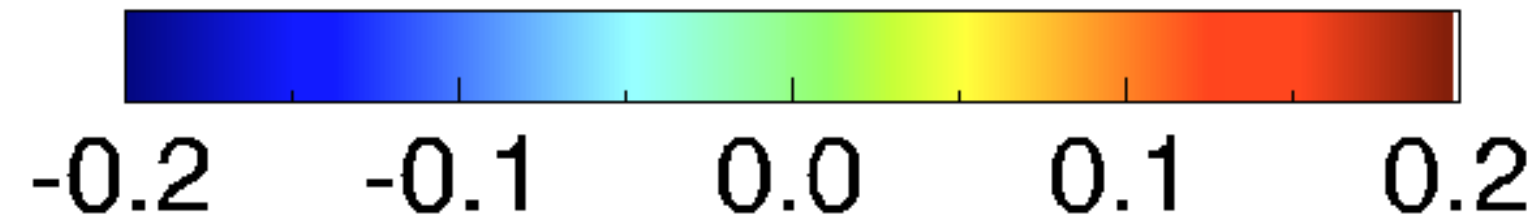}
\caption{Poynting flux on the bottom plate in the middle of the first cycle (a) for each simulation, and during the last cycles for the $p=100\%$ (b), $p=50\%$ (c), and $p=0\%$ (d) preferences. }
\label{fig:poynting}
\end{figure*}

Figure \ref{fig:poynting} shows the Poynting flux density $S_n$ on the
bottom plate at the peaks of the first and last cycles for each of the
three cases. The flux depends upon the stress exerted by the field on
the surface, which for our case varies only with the horizontal field
component $\vecB_h$, because the vertical field is constant. Slow
twisting of field lines by the convective flow $\vecv_h$ does work on
the field, injecting energy at a rate determined by $\vecv_h \cdot
\vecB_h$. This stress can be partially relieved by the rearrangement
of field lines due to reconnection in the overlying corona. As
expected, the Poynting flux density during the first cycle is
identical for the three cases, and corresponds exactly to the
rotational cell pattern at the photosphere
(Fig.~\ref{fig:poynting}a). Early on, before the onset of any
reconnection, the only effect of the motions is to inject twist
(horizontal) flux into each flux tube driven by a rotational cell. The
energy associated with this twist flux is independent of the sign of
the twist (clockwise or counter-clockwise); consequently, the Poynting
flux is everywhere positive, and is essentially identical in each
rotation cell. In principle, there is some variation due to the fact that
the twisted flux tubes at the periphery of the driving pattern can
expand more than those in the interior, but this effect is extremely
small. \par

The situation is very different later in the simulation. The field
acquires a complex distribution of
horizontal flux, and the magnetic stress at the photosphere can either
oppose or reinforce the motion. If it opposes, then the flow does work
on the field and there is a positive injection of magnetic energy. If
the stress and the motion are in the same direction, then energy is
removed from the field as it de-stresses, and the Poynting flux density
is negative. The field, of course, also de-stresses due to reconnection
in the volume. \par

Panels (b), (c), and (d) in Figure \ref{fig:poynting} show the Poynting flux
in the middle of the last cycle for the three helicity injection
preferences. As discussed above, the flux density has both positive
and negative regions, but a preponderance of positive. We note
that the maximum flux is a factor of 3 or so greater than that halfway
through the first cycle. Since the velocities at the photosphere
have fixed magnitude, this maximum sets a limit on the maximum horizontal
field there. In particular, the horizontal field does not exceed that
induced by two full cycles of rotation. This limit indicates that
reconnection is highly efficient at keeping the field not far from its
potential state. We found in previous work \citep{Knizhnik15,Zhao15}  
that for the 100\% case, reconnection sets in well before the flux
tubes acquire two turns of twist. For the 0\% case, two full twists are
generally sufficient to induce kinking of the flux tubes, which again
results in efficient reconnection. In any event, since the whole flow
pattern is rotated randomly after each cycle, current sheets and
reconnection will inevitably appear during the second cycle,
irrespective of the helicity preference. \par

There are a number of important features in the Poynting flux
maps. First we note that, as with the temperature, the three cases
show marked differences at the boundary. For the 0\% case, there is no
significant difference between the fluxes at the center versus the
boundary; the pattern appears to be roughly random everywhere. The
100\% case, on the other hand, shows distinct structure at the
boundary, particularly in the last row of photospheric rotation cells. 
At the outer edge of each of these cells, the flux is strongly positive,
whereas it is negative at the inner edge. It is straightforward to
understand this pattern. At the outer edge, the flow is in the
direction that increases the large-scale shear of the filament channel field;
consequently, it injects a large, positive Poynting flux there. At the
inner edge, in contrast, the flow acts to decrease the shear, resulting
in a negative Poynting flux. We emphasize that the buildup of large-scale 
shear is \textit{not} due to this systematic pattern in the outermost
cells. Each cell acts only to impart a
localized rotation of its corresponding flux. In other words, a cell
at the boundary does not increase the large-scale shear, it only
produces a localized twist in the shear flux. If there were no
reconnection, no large-scale shear would ever build up. The
reconnection causes the small-scale twist flux to cancel in the
interior and to build up systematically at the boundary.

The 50\% case is intermediate to
the other two. There is mainly strong positive flux at the outer edge,
but since $1/3$ of the rotations are counter-clockwise, there is also
some strong negative Poynting flux. Note that even if we ran this case
for twice as long as the 100\% case, so that the same amount of shear
flux builds up at the boundary, the 50\% case would still exhibit a
mixture of positive and negative Poynting flux. \par

Figure \ref{fig:poynting} shows that there are also clear differences
between the three cases in the interior of the driven region.  As with
the temperature (Fig.~\ref{fig:temperature}), the 100\% case shows the
least structure. All the interior cells appear to be injecting a
primarily positive energy flux of intermediate magnitude. In fact, it
is surprising how little negative flux is present in the map,
given that the whole pattern rotates randomly after every cycle, so
that some of the motions should be de-stressing the field. The other
two cases exhibit distinct regions of negative Poynting flux, but
again primarily positive. Even for the 0\% case, where each cell has a random
sense of rotation, the Poynting fluxes are primarily positive. This result is
critically important, because a net positive flux is required for
continued coronal heating. \par

The results in Figure \ref{fig:poynting} can be explained by the
detailed structure of the magnetic field. As discussed above,
reconnection is very efficient at preventing the field from
deviating far from its potential state. When the reconnection
is pervasive, the Poynting maps should resemble that
of panel (a), and the flux density should be positive everywhere. 
As shown in KAD17, the 100\% helicity injection case produces a
magnetic field that has the least amount of small-scale structure, and
is closest to the potential state. Consequently, the interior region
of the Poynting map in panel (b) is quite similar to that of panel (a):
there is weak positive flux everywhere (note the scale change between 
these two panels). For the other two cases, in contrast, the
field has some localized regions of strong non-potentiality, so
the Poynting maps show locations of strong negative and positive 
flux. For the most part, nevertheless, the field is sufficiently near
potential that the Poynting flux density is weakly positive overall. \par

In order to understand the effect of spatial variations of the
Poynting maps on coronal heating, we calculated the total energy
injected into the domain.  Area integrations of the flux density $S_n$
over the top and bottom boundaries ($x=0,1$) yield the total Poynting fluxes
$P_n$ of electromagnetic energy per unit time,
\beg{IntPoynting}
P_n(t) = \oint_S{S_n(x,y,z,t) dA},
\done 
which are shown in Figure
\ref{fig:energy}a. The dominant temporal behavior of the total Poynting flux
is sinusoidal, as determined by the oscillatory nature of the convective
cells described in \S \ref{sec:model}. We note that, for the most part,
the magnitude of the peak flux varies with the helicity
preference. The 100\% case shows the highest peak in most cycles,
followed by the 50\% case, and then the 0\%. The actual magnitudes,
however, vary significantly from cycle to cycle, and there are even
cycles where the 0\% case exhibits the highest peak. This result
demonstrates the random nature of the driving and the reconnection
response. Presumably, we would obtain fewer such fluctuations if we
employed a much larger number of driving cells than the 61 used in these
simulations, but the temporally averaged results would be similar. 

Another important feature of Figure \ref{fig:energy}a is that the peak
flux does not show much of an increase after the first cycle or
so. This again implies that the field never deviates much
from its initial potential state, except at the ``PIL.''  To quantify
the long-term
evolution of the system, we plot the cycle-averaged values
$\langle P_n \rangle$ in Figure \ref{fig:energy}b. After about
five cycles, the $p=0\%$ case appears to have reached a statistically steady state
in which the energy injected is approximately constant from cycle to
cycle, albeit with substantial fluctuations. The average flux over the
simulation is
\beg{PF0} \langle P_n
\rangle^s_{0\%} \approx 5.4 \times 10^{-2}, \done
where the superscript $s$ indicates the quasi-steady value.  The
associated average Poynting flux density is
\beg{PFD0} \langle S_n
\rangle^s_{0\%} \approx 9.0 \times 10^{-3}.  \done
The fact that this
case achieves a steady state is not surprising. Eventually, the small-scale
magnetic stresses in the corona  must build up to the point where further
tangling simply results in faster energy release by reconnection, leading
to a steady state, exactly as predicted by \citet{Parker72} in his seminal
work. What is surprising, perhaps, is how quickly this state is
achieved, after only a few cycles of driving. As argued by Parker
and by many others \citep[e.g.,][]{vanBallegooijen85}, random footpoint
motions are highly efficient at forming current sheets in three
dimensions. \par

As is to be expected, the cases with net helicity injection do not 
reach a steady state; instead, they exhibit a weak quasi-linear growth
at later times. The origin of this trend is the constant increase of
the global shear field (the filament channel) at the system boundary,
which continuously increases the free energy in the system. This
shear field exerts a systematic stress at the photosphere that
increases monotonically as the shear field increases. Also as expected,
the effect of the global shear field is stronger in the 100\% case
than in the 50\%. To determine this effect more accurately, we plot in
Figure \ref{fig:energy}c the time-integrated Poynting flux, i.e.,
the total energy injected into the system, for the three
cases. Note that the 0\% case shows a linear increase with time,
whereas the other two curve upward, indicating that the
Poynting fluxes for these cases is not constant. Furthermore, the 100\%
curve shows a stronger upward curvature than the 50\% case. \par


\subsection{Magnetic Energy}\label{sec:ME}
In order to determine the effect of the photospheric driving motions on the coronal
energetics, we calculated the evolution of the total magnetic and plasma
energies. The magnetic (free) energy added to the volume is
\beg{free_energy}
\Delta M(t)=\frac{1}{8\pi}\int_V{\left[ B^2\left(x,y,z,t\right)-B^2\left(x,y,z,0\right) \right] dV}
\done
and is displayed in Figure \ref{fig:energy}d. The oscillatory pattern
in each curve is due to the sinusoidal temporal profile of the driving
motions. Clearly, the magnetic energy depends strongly on the helicity
preference. For the 0\% case, the free energy levels off after a dozen
cycles or so to a fairly low level that corresponds to $\approx 30$\%
of the initial potential energy. The conclusion is that, after the
initial magnetic energy rise, all the Poynting flux is converted to
plasma thermal energy. \par  

The two cases with a helicity preference exhibit a clear continued rise in
magnetic free energy.  The $p=100\%$ case, for example, has more than
triple the free energy of the $p=0\%$ case, and almost double that of
the $p=50\%$ case. The differences in the free energies, and the
relatively large free energy of the 100\% case, are due primarily to the
highly sheared filament channel that forms at the boundary of the
driven region. This can be seen from the following analysis. \par

For $p=100\%$, the magnetic free energy in the corona at the end of
the simulation can be estimated by approximating the final
configuration as a uniform vertical field everywhere except for a band
of global shear flux encircling the outer boundary. The free energy contained in
the volume is dominated by the shear field at this boundary, since it
is the only location where the magnetic field is highly
nonpotential. With this approximation, the free energy of the final
configuration for the 100\% preference, after 20 cycles, is given by
\beg{approxfreeenergy}
\Delta M_{100\%}^{20} \approx \frac{B_{\phi}^2}{8\pi}V_b.
\done
Here $V_b$ is the volume occupied
by the band of twist field, whose strength is $B_{\phi}$. The band is
centered at distance $a_b$ from the central axis, its width is $w$,
and its height is $L_x$, so its volume is
\beg{VPIL}
V_b \approx 2\pi a_b w L_x.
\done

The twist field strength is given by
\beg{twistfield}
B_{\phi} = \frac{\Phi_{tw}}{wL_x},
\done
where $\Phi_{tw}$ is the twist flux through a vertical plane. In KAD15 we
related the twist flux to the helicity injected into the corona, so
that, at each instant, $\Phi_{tw}$ is known.  Substituting Eqs.\
\ref{VPIL} and \ref{twistfield} into Eq.\ \ref{approxfreeenergy}
yields
\beg{approxfreeenergy2}
\Delta M_{100\%}^{20} \approx=\frac{a_b}{4wL_x}\Phi_{tw}^2.
\done
Inserting the domain height $L_x=1$ and the empirical simulation
values $w=0.3$ and $a_b=1.0$, we find

\beg{Wpredict}
\Delta
M_{100\%}^{20} \approx 8.3\times10^{-1} \Phi_{tw}^2.
\done
For $p=100\%$, KAD15 predicted and confirmed numerically that the
twist flux $\Phi_{tw}$ depends only upon the amount of helicity
injected per convection cell, which for this case satisfies
\beg{predphitwform}
\Phi_{tw}=1.7\times10^{-2} \left[ t - \frac{\tau}{2\pi} \sin \left( 2\pi\frac{t}{\tau} \right) \right].
\done
After $20$ cycles, $t=20\tau$, we find
\beg{predphitw}
\Phi_{tw}=1.16.
\done
Using this value in Eq.\ \ref{Wpredict} yields finally
\beg{Wfreepredictval}
\Delta M_{100\%}^{20} \approx 1.12.
\done

This value agrees quite well with the $p=100\%$ (orange) curve in
Figure \ref{fig:energy}d. Note that it also is close to the initial
magnetic energy in the volume, $M(0) = 1.125$, indicating that our
approximation is starting to break down. Once the shear energy becomes
of order the initial energy, then the presence of the shear will produce
significant displacements in the initially uniform vertical field,
thereby increasing the energy in that component as well. This is evident from
Figure \ref{fig:temperature}b. Furthermore, our shear field has built up to ``too high'' a
value compared to actual coronal fields. Both theory and observations
of CMEs and coronal jets have consistently found that eruption invariably
occurs when the free energy reaches 50\% or so of the initial potential
energy \citep[e.g.,][]{Antiochos99, Karpen12, Wyper17}. \par

\begin{figure}[!p]
\centering\includegraphics[scale=0.5, trim=0.0cm 12.5cm 0.0cm 0.0cm]{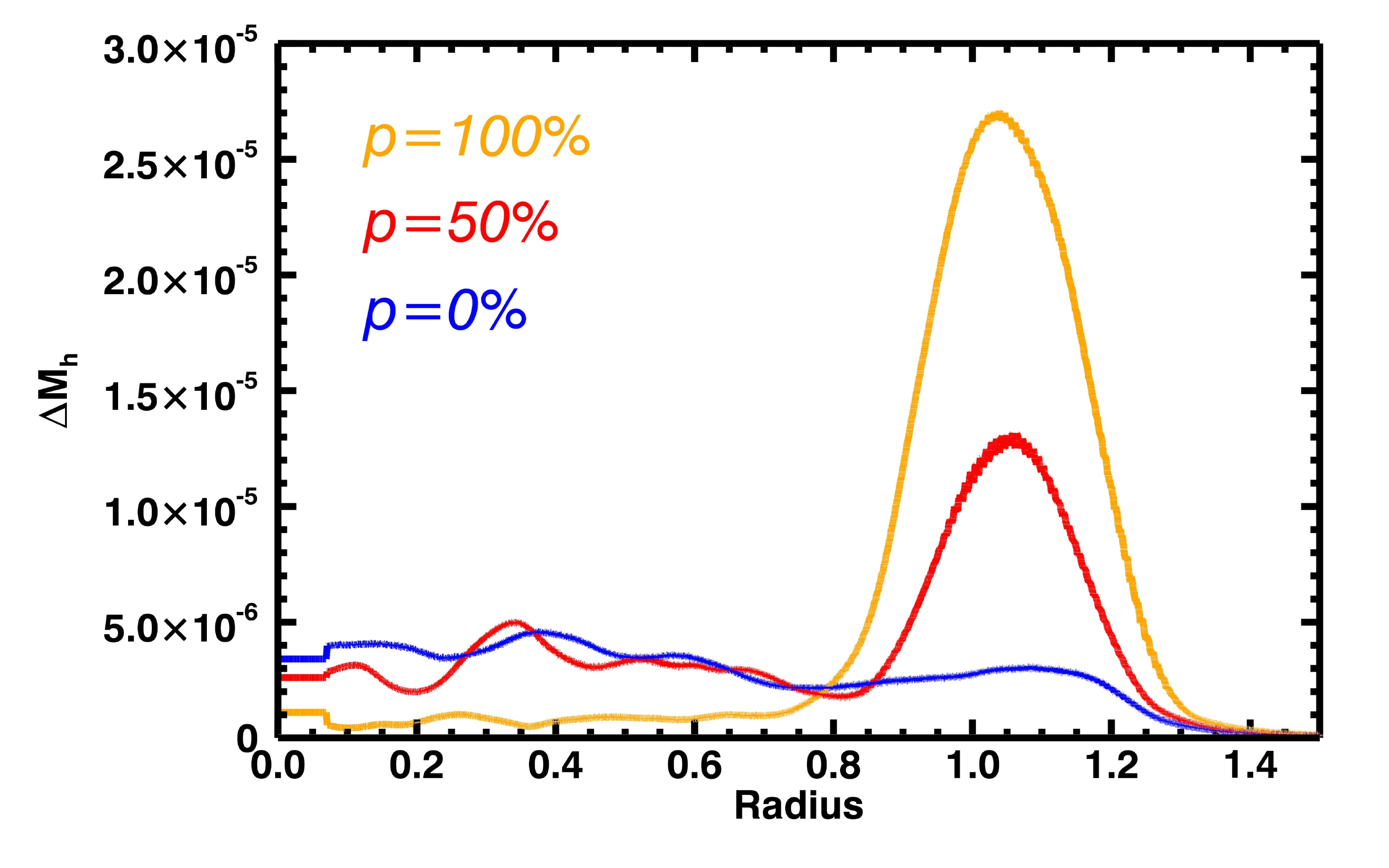}
\vspace{60mm}
\caption{Free magnetic energy density, integrated vertically and azimuthally, vs.\ radius at the end of the simulation for the $p=100\%$ (orange), $p=50\%$ (red), and $p=0\%$ (blue) preferences. (cf.\ Eq.\ \ref{Wr}). }
\label{fig:Wr}
\end{figure}

As further evidence that the free energy in our system is due
primarily to the buildup of the filament channel, we plot in Figure
\ref{fig:Wr} the quantity
\beg{Wr}
\Delta M_h(r,t) =\frac{1}{8\pi}\int_0^{2\pi}{d\phi}{\int_0^{L_x}\vecB_h^2(x,r,\phi,t)dx}
\done
at the end of the simulation for each of the three cases. This
quantity is simply the energy density of the horizontal field integrated
over height and azimuth for cylindrical shell of radius $r$. Its
distribution shows precisely where the magnetic free energy is stored
as a function of distance from the center of the region. We neglect
the contribution from the vertical field, which contains minimal free
energy except near the very end of the 100\% case. Especially for the
$p=100\%$ case, but also for the $p=50\%$ case, most of the free
energy is stored near the boundary of the driven region. In sharp
contrast, for $p=0\%$ the magnetic free energy is spread relatively
uniformly throughout the driven region, and it is much smaller than
the peak values obtained for $p=100\%$ and $p=50\%$. It is tempting to
infer from our results that most of the free energy in the solar
corona is stored in filament channels, but some caution needs to be
taken here due to the geometrical differences between the corona and
our closed-box domain. Observations clearly show that the strong
non-potentiality of the coronal field (i.e., any electric current) is
highly concentrated in the filament channel, as in our
simulation. However, since the corona is an infinite domain in which
the field generally drops off rapidly with height, the presence of a
filament channel causes large upward displacements and increases the
free energy of the overlying flux. This is inherent in all CME
models: the extra magnetic pressure of the filament field must
be balanced by an increased magnetic tension in the overlying flux
(Antiochos et al 1999). In our system, the side and top walls of the
simulation box can exert confining forces, but, of course, there are
no walls in the corona. \par


\subsection{Internal Energy}\label{sec:IE}

\begin{figure*}
\begin{minipage}[b]{0.5\linewidth}
\centering\includegraphics[scale=0.7, trim=5.0cm 0.0cm 0.0cm 0.0cm]{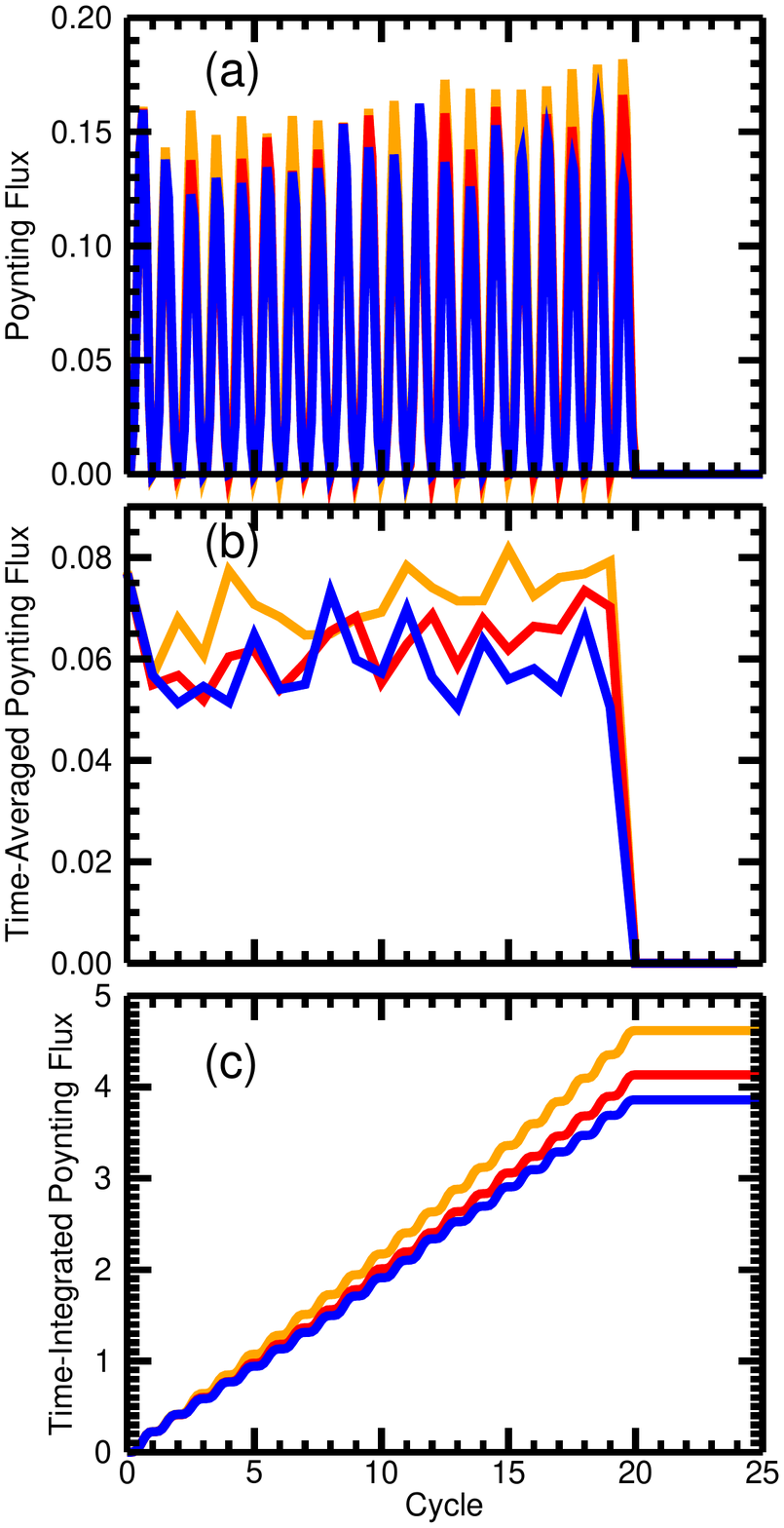}
\end{minipage}%
\hfill
\begin{minipage}[b]{0.5\linewidth}
\centering\includegraphics[scale=0.7, trim=5.0cm 0.0cm 0.0cm 0.0cm]{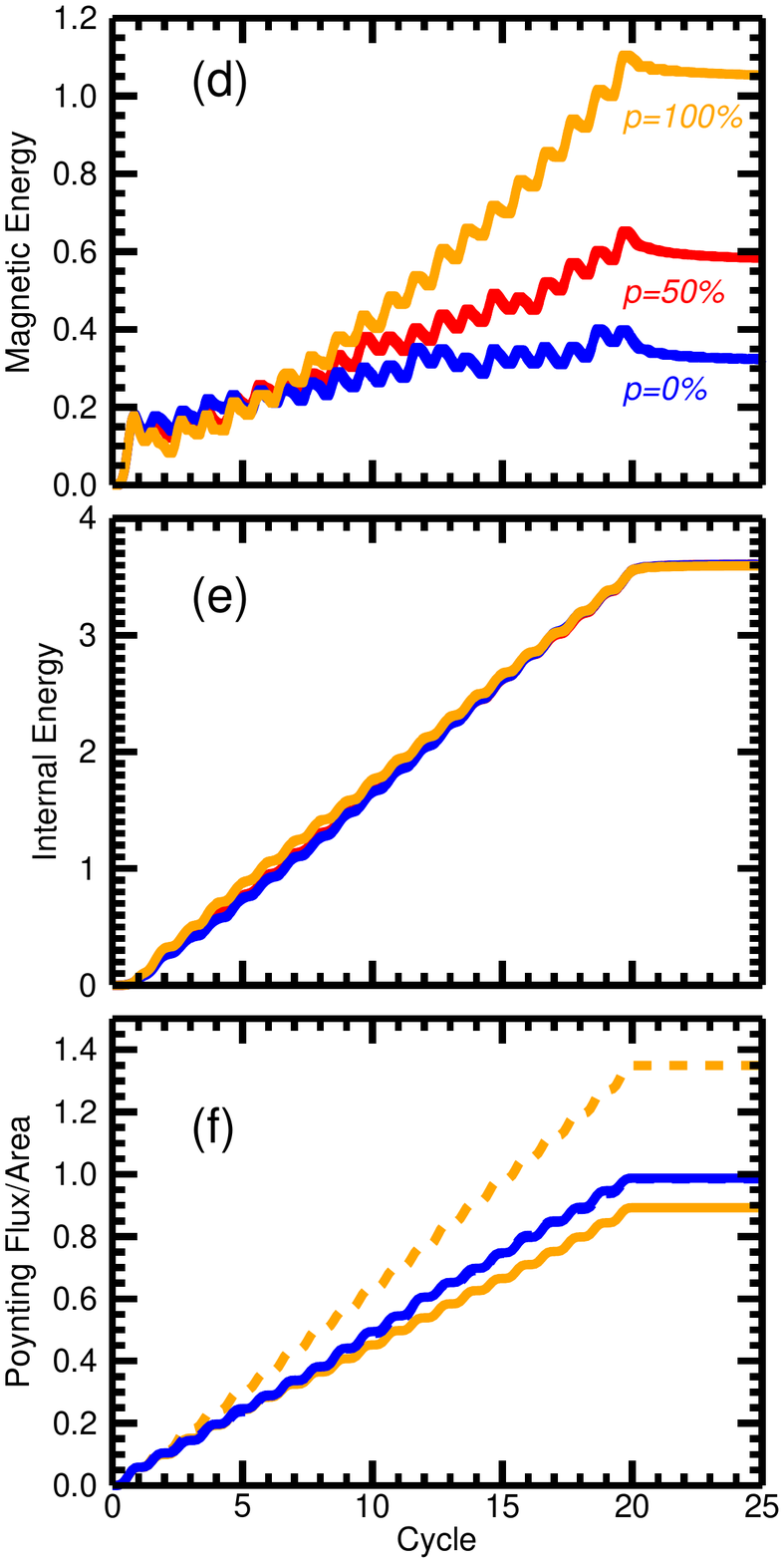}
\end{minipage}
\vspace{0mm}
\caption{Left: Instantaneous (a), cycle-averaged (b), and time-integrated (c) Poynting fluxes vs.\ time for the $p=100\%$ (orange), $p=50\%$ (red), and $p=0\%$ (blue) preferences. Right: Corresponding free magnetic (d), and excess internal (e) energy. In (f), we plot the Time-Integrated Poynting flux per unit area inside disks of radii $0.25<r<0.75$ (solid curves) and $0.75<r<1.00$ (dashed curves) for the $p=100\%$ and $p=0\%$ cases. {\color{black} Note that the dashed blue line is completely covered by the solid blue line.} }
\label{fig:energy}
\end{figure*}

The results above demonstrate that the helicity preference of the
driving motions has a strong effect on the magnetic energy in the
corona. It would seem likely, therefore, that it should also have a
similar effect on the plasma energy, but this is not the case. The
increase in internal energy in the volume is given by
\beg{internal_energy}
\Delta I(t)=\int_V{\left[
    U\left(x,y,z,t\right)-U\left(x,y,z,0\right) \right] dV}
\done
and is displayed in Figure \ref{fig:energy}e. It accounts for more than
$75\%$ of the total energy added to the volume even for the 100\%
helicity preference. Clearly, the majority of the injected energy
is converted into heat via magnetic reconnection, rather than remaining
stored in the magnetic field. The important result of our study,
evident in Figure \ref{fig:energy}e, is that \emph{the amount of
  heating is only weakly dependent on the helicity preference}. At
least for these particular simulations, the increase in internal
energy appears to be almost the same for $p=100\%$, $p=50\%$, and
$p=0\%$, especially near the end of the simulations. \par

This result is somewhat surprising given that the photospheric
Poynting fluxes for the three cases show measurable differences. 
Concentrating on the 100\% and 0\% cases, we note from Figures
\ref{fig:energy}a and \ref{fig:energy}b that the Poynting flux
for the former case is noticeably higher throughout the evolution
than for the latter. In contrast, between about cycles 5 to 12 the
internal energy for the 100\% case in Figure \ref{fig:energy}e is
only slightly higher than that for the
0\%, and the latter catches up by about cycle 16. Over the last 4
cycles of the three simulations, their internal energies are
virtually indistinguishable. It seems unlikely that if we were to
continue the driving for more cycles beyond 20, a clear separation
of the internal energies would appear. Moreover, such a result
would be difficult to relate to the corona, because the simulated
plasma beta has already risen well above unity by cycle 20. \par

Although the degree of agreement between the internal energies of the
three cases is surprising, the basic result that the heating rate is
weakly sensitive to the helicity injection is not unexpected
\citep[e.g.,][]{WS11}. For turbulent 3D MHD systems, both experiment
and theory have shown that free energy and helicity undergo very
different evolutionary tracks: the energy cascades down to the
dissipation scale where it is converted to thermal energy, while the
helicity cascades up in scale and simply collects there 
\citep[e.g.,][]{Biskamp93}.
Of course, our system is not truly turbulent. The line-tying
at the photosphere and the low plasma beta restrict the corona from
undergoing an actual turbulent cascade wherein eddies of continuously
decreasing scale are spontaneously generated by large-scale
driving. The corona, however, does not require turbulence for its
heating, only ubiquitous reconnection, and this is exactly what the
line-tying and the low beta produce. \par

\emph{A priori}, it would seem likely that the free-energy evolution
of the three cases would be quite similar. The
energy is injected by small-scale quasi-random motions with the same
scale, speed, etc.\ and, except for the filament channel, the coronal
field remains rather close to the potential state. The
differences in the heating, therefore, are likely to be small. If
anything, we would expect the 100\% case to have the lowest heating,
because the driving is such that it is maximally efficient at
inducing reconnection. In other words, this case has the largest
effective resistivity and, as argued many years ago by \citet{Parker72},
a larger resistivity results in a smaller heating rate. \par

To investigate this hypothesis quantitatively, we plot in Figure
\ref{fig:energy}f the time-integrated Poynting flux through the
boundaries, analogous to Figure \ref{fig:energy}c, but for subregions on
the boundaries. In order to make direct comparison easier, we plot the
flux per unit area and, to reduce clutter, we include only the 0\% (blue)
and 100\% (orange) cases. The {\color{black} dashed} curves represent the region
$0.75 < r < 1.0$, which corresponds to the filament channel, and
the {\color{black} solid} curves represent the main interior region $0.25 < r < 0.75$.
We omit the exact center because it is somewhat artificial, in that
the twist there is never displaced by the random rotation of the pattern. \par

The first result to note from Figure \ref{fig:energy}f is that the
integrated flux per area in the filament channel region is much larger
for the 100\% case than for the 0\%. This is to be expected from
comparison of Figures \ref{fig:poynting}b and \ref{fig:poynting}d. The
rotations in the outermost ring for the 100\% case show a substantially
larger positive Poynting flux than negative, especially for those
rotations along the sides of the hexagonal pattern. These rotations
intersect the boundary of the filament channel region, so the flow
along their outer edges moves in the direction of the shear and
provides a strong positive Poynting flux, while their inner edges
are in the weak-shear zone and produce only a weak negative. For the 0\%
case, on the other hand, the Poynting flux is randomly distributed
along each flow irrespective of the flow's location. The enhanced
Poynting flux in the vicinity of the filament channel for the 100\%
case provides the
energy for the shear field there, and indirectly enhances the internal
energy throughout the volume. Recall that the temperature in the
interior is systematically larger in this case due to the
``squeezing'' of the interior by the band of strong shear flux. We
conclude that if the only driving were in the outer ring, $0.75 < r <
1.0$, the internal energy of the 100\% case would be larger than that
of the 0\%. \par

\begin{figure}[!p]
\centering\includegraphics[scale=0.4,trim=0.0cm 4.0cm 1.0cm 1.0cm, clip=true]{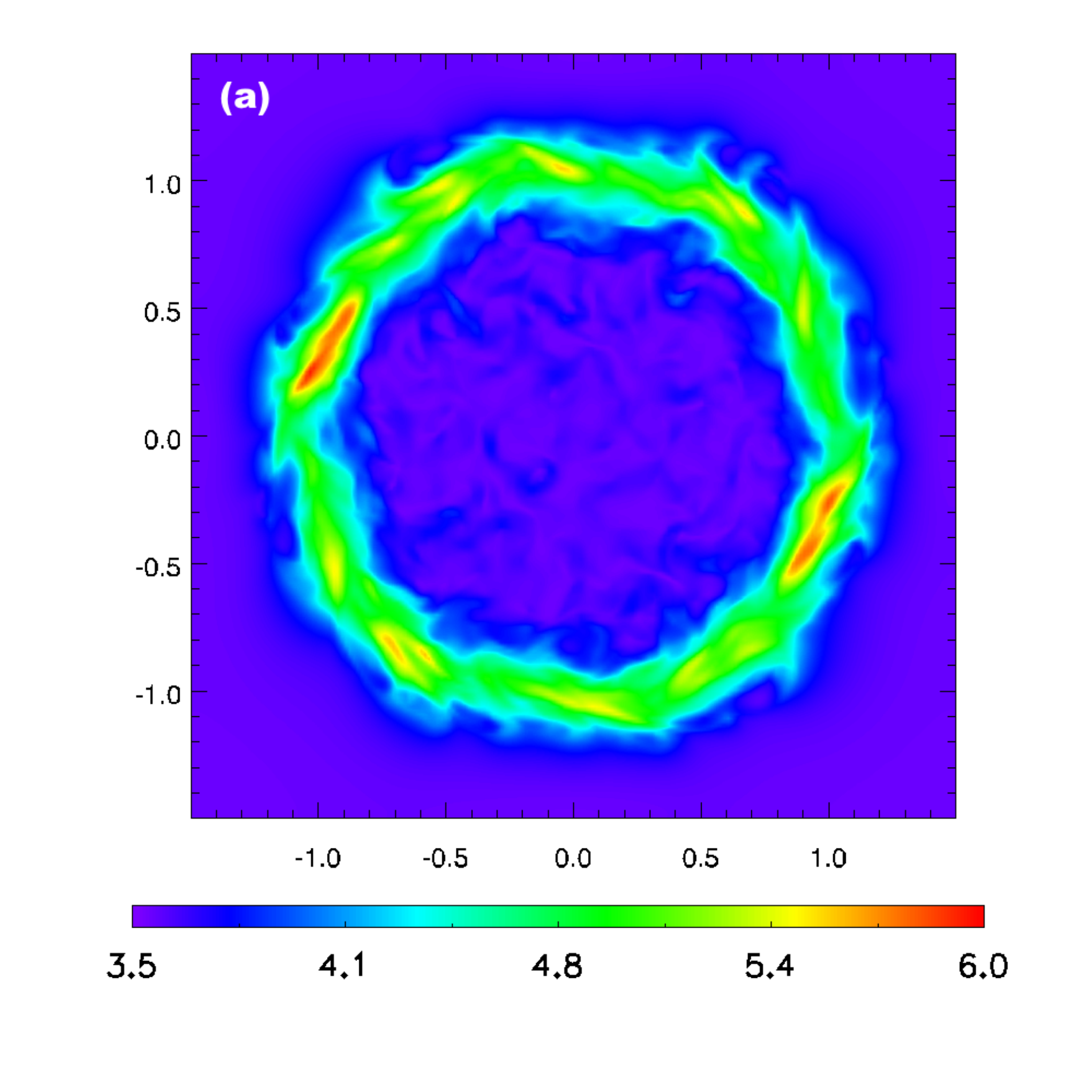}
\centering\includegraphics[scale=0.4,trim=0.0cm 4.0cm 1.0cm 1.0cm, clip=true]{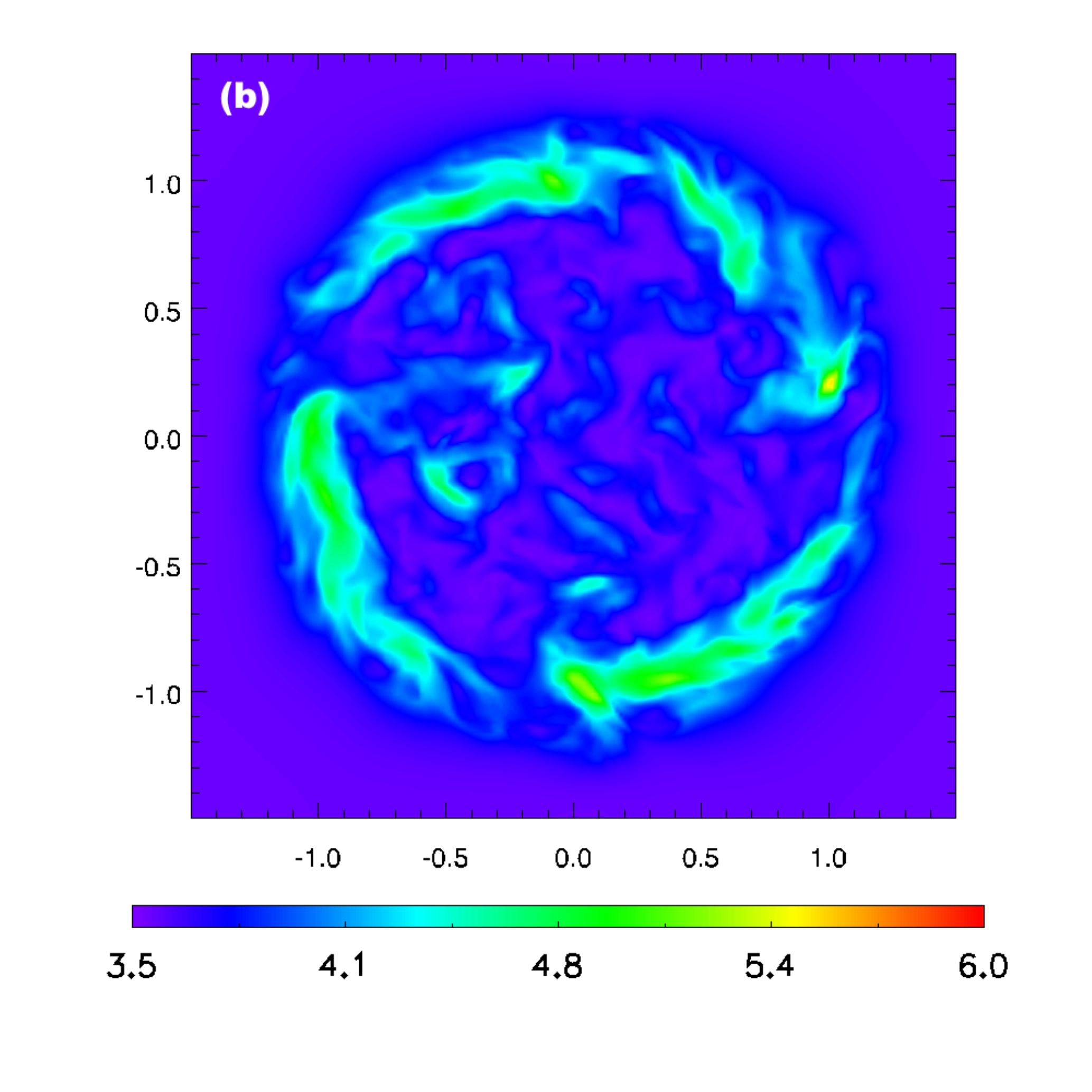}
\centering\includegraphics[scale=0.4,trim=0.0cm 0.0cm 1.0cm 1.0cm, clip=true]{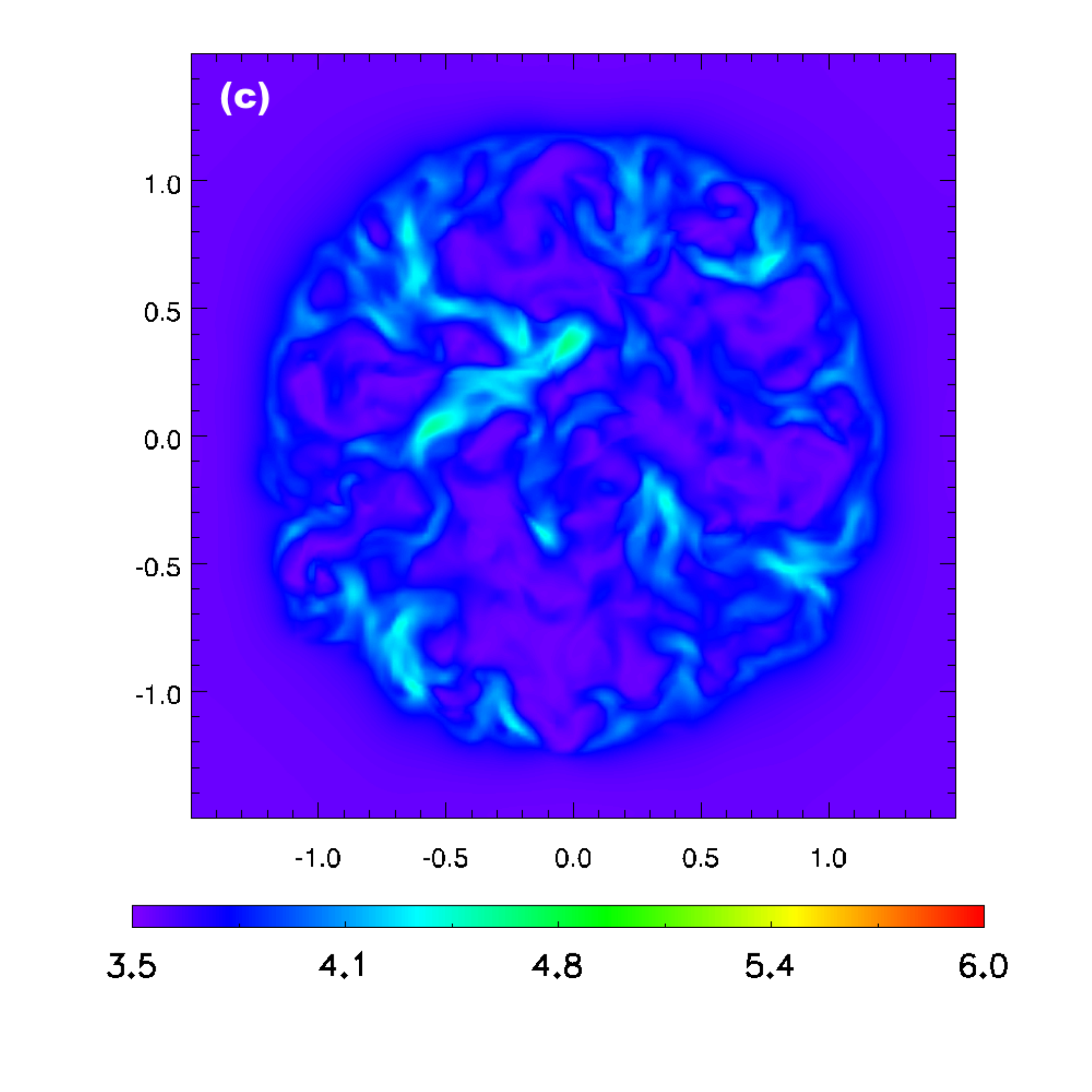}
\caption{$|B|$ at the bottom boundary at the end of the simulation for the (a) $p=100\%$, (b) $p=50\%$ and (c) $p=0\%$ preferences. }
\label{fig:bmag}
\end{figure}

The driving in the interior region, however, has the opposite
effect. Figure \ref{fig:energy}f shows that in the region
$0.25 < r < 0.75$, the Poynting flux per area is significantly larger
for the 0\% case than the 100\%. The reason is simply the lack of
large-scale twisting and tangling that develop for 100\% helicity
injection, as discussed in KAD17. Figure \ref{fig:bmag} shows the
variations in the magnitude of the horizontal magnetic field,
\beg{mag_horizontal}
B_h = \sqrt{B_y^2 + B_z^2},
\done
at the bottom boundary for the three
cases. We actually plot the total $B$ but, since the vertical
component stays uniform throughout the simulations, the contours in
Figure \ref{fig:bmag} reflect $B_h$. The 100\% case exhibits the
strong $B_h$ ring of the filament channel, but almost no $B_h$ structure
in the interior. Reconnection is amazingly efficient at ``transporting''
all the helicity to the PIL, leaving a very smooth coronal-loop
region. The 0\% case, in sharp contrast, contains clumps of large $B_h$
located randomly throughout the driven region. Since the Poynting flux
is directly proportional to $B_h$, the flux is noticeably larger for
the smaller helicity preference. \par

In applying these results to the observed corona, a number of effects
must be kept in mind. First, the ratio of the areas of
the filament channel versus coronal-loop regions will vary, thereby
changing the cumulative agreement evident in Figure \ref{fig:energy}e. For
example, if we were to double the radius of our system, but keep the
size of each photospheric rotation the same, then the area of the
filament channel would double, but the area of the interior (coronal
loop) region, along with the number of rotations, would increase by
a factor of four. In this case, a large helicity injection preference
would result in a cooler corona with less internal energy. 
Another effect is that magnetic flux in the corona
can expand outward and, hence, change the internal energy of the
plasma. The buildup of the filament channel will cause the overlying
coronal-loop flux to expand upward, thereby decreasing $B_h$ at the
photosphere and the resulting Poynting flux. \par

The final important effect is that of numerical resolution. For our
simulations, in which the effective resistivity is numerical, the scale
of the grid determines the dissipation scale. As we increase the
refinement level of the simulations, we expect the effective
resistivity to decrease and, consequently, the heating to
increase. This will likely enhance any differences in the heating
rates between the different cases. The scaling of the heating with
numerical resolution is also a critical issue for applying our results
quantitatively to the observed corona. Therefore, in the following
section we discuss in detail the effect of numerical resolution. \par


\section{Grid-Resolution Study}\label{sec:resolution}
The key question for our study is, how do the heating rates in our
calculations compare to the observed coronal heating? The main challenge
with addressing this question using any simulation is that the
numerical Lundquist numbers are inevitably far lower than those in the
corona. Consequently, the best that can be done is to perform a scaling
study and attempt to extrapolate to coronal parameters. In our
calculations, the conversion of injected magnetic energy into plasma
heating and internal energy occurs via magnetic reconnection across
current sheets formed at the numerical grid scale; therefore, in order
to scale on our results with effective Lundquist number, we performed
simulations at several different grid resolutions. To avoid the
complication of contributions from the filament channels, we elected
to conduct this study for the $p=0\%$ case. We used four different
levels of refinement, denoted by $l$ = 1, 2, 3, and 4, corresponding
to $8$, $16$, $32$, and $64$ grid points across each rotation cell,
respectively. The results presented in \S \ref{sec:results} were
obtained with $l=3$, or $32$ grid points across each cell. \par

\begin{figure*}
\begin{minipage}[b]{0.5\linewidth}
\centering\includegraphics[scale=0.7, trim=5.0cm 0.0cm 0.0cm 0.0cm]{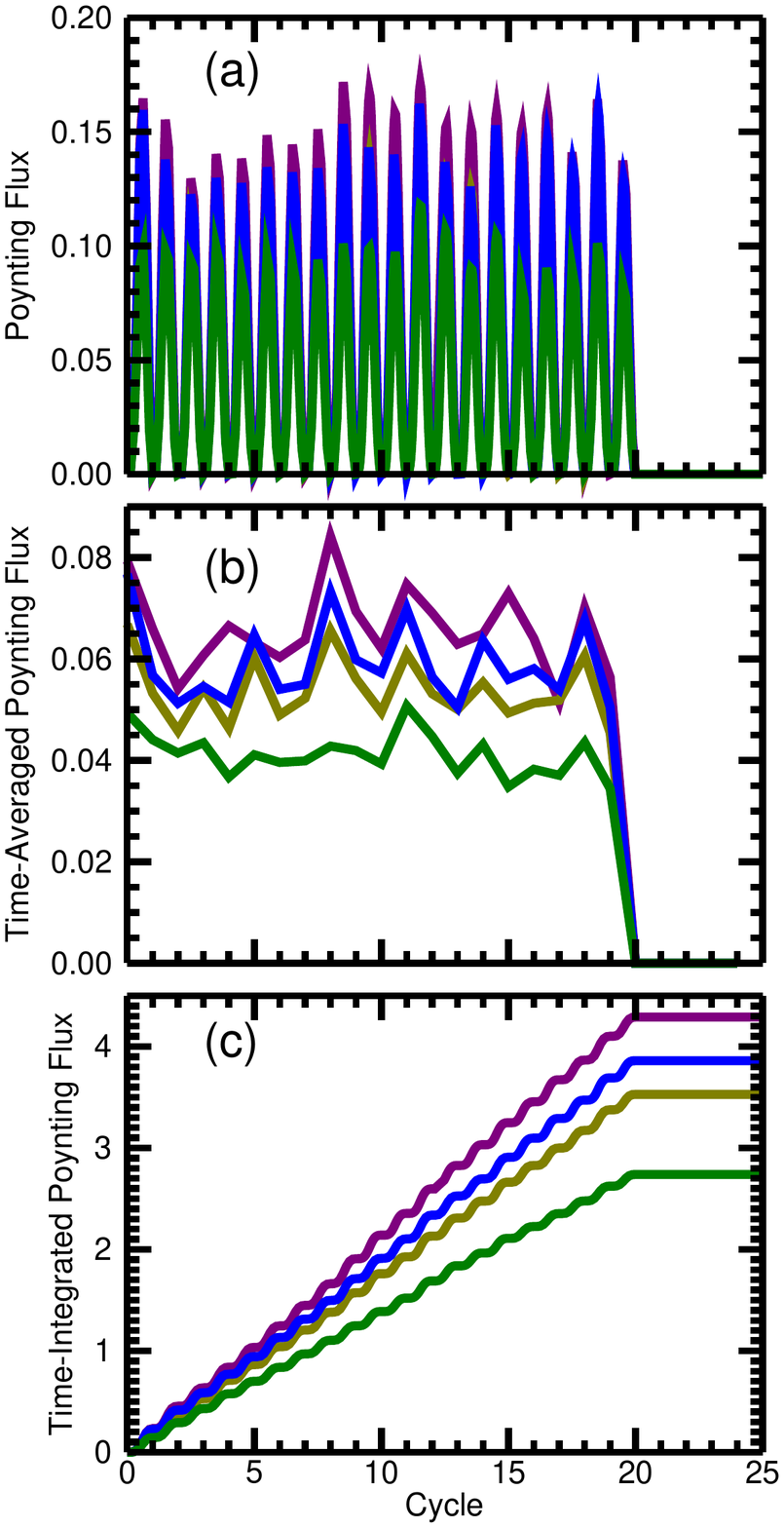}
\end{minipage}%
\hfill
\begin{minipage}[b]{0.5\linewidth}
\centering\includegraphics[scale=0.7, trim=5.0cm 0.0cm 0.0cm 0.0cm,clip=true]{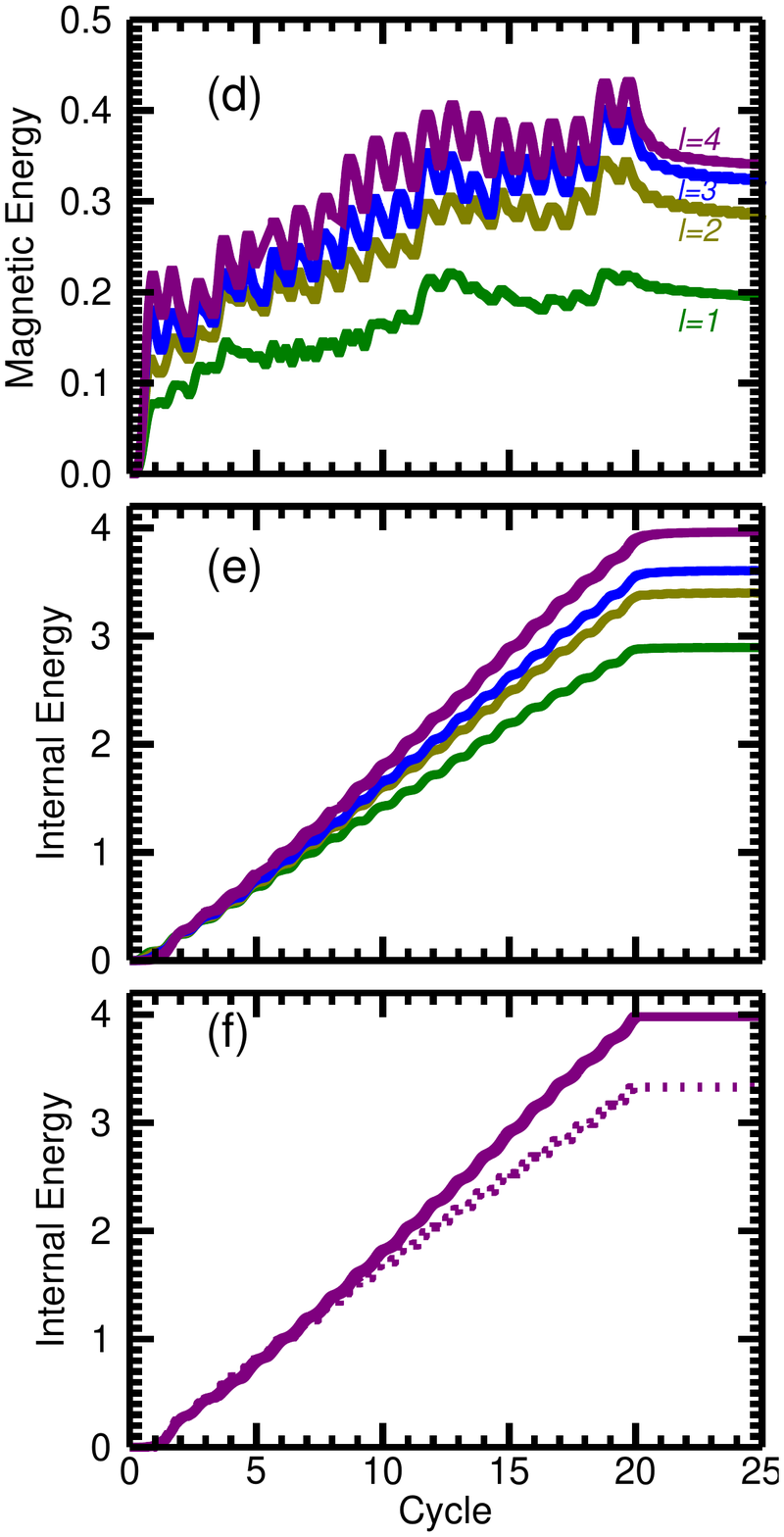}
\end{minipage}
\vspace{0mm}
\caption{Left: Instantaneous (a), cycle-averaged (b), and
  time-integrated (c) Poynting fluxes vs.\ time for the $p=0\%$
  preference for simulations at refinement levels $l=1$ (green), $l=2$
  (olive), $l=3$ (blue), and $l=4$ (purple). Right: Corresponding free
  magnetic (d), excess internal (e), and excess total energy. (f) shows the excess internal energy for the $l=4$ $0\%$ (solid) and $100\%$ (dashed) helicity cases. }
\label{fig:energy_ref}
\end{figure*}

The first five panels (a-e) of Figure \ref{fig:energy_ref} display the same
quantities as the corresponding panels in 
Figure \ref{fig:energy}, but for fixed $p=0\%$ at the different
refinement levels $l=1,...,4$. In general, all quantities increase as
the refinement level increases. The instantaneous, time-averaged, and
time-integrated Poynting fluxes (Fig.~\ref{fig:energy_ref}a-c)
exhibit increasing amplitudes and fluctuation levels, reflecting the
enhanced small-scale structure and dynamics that follow from the
improvements in resolution. These features also are evident in the
plots of the magnetic and internal energies (Fig.~\ref{fig:energy_ref}d-e). 
In contrast to our finding in \S
\ref{sec:results} that the internal energy change, and thus the
heating rate, is only weakly dependent on helicity preference $p$,
here we observe that these quantities are quite dependent upon the
refinement level $l$, increasing monotonically as the grid resolution
increases. During the first five cycles of rotation, the curves all
rise together at essentially the same pace; thereafter, however, they
begin to diverge and adopt slightly different slopes, i.e., heating
rates. \par

The last panel (f) of Figure \ref{fig:energy_ref} shows the evolution
of the internal energies for the 0\% (solid) and the 100\% (dashed) cases
at the highest numerical resolution.  We now see a clear deviation
between the two, with the 100\% helicity preference resulting in a
lower heating rate, as expected. The difference is not substantial,
however, only of order 15\% or so. We conclude that the helicity
preference, hence the \emph{net} helicity injection, plays only a
minor role in determining the magnitude of the coronal heating. \par

\begin{figure*}[!p]
\centering\includegraphics[scale=0.5, trim=0.0cm 0.0cm 0.0cm 0.0cm]{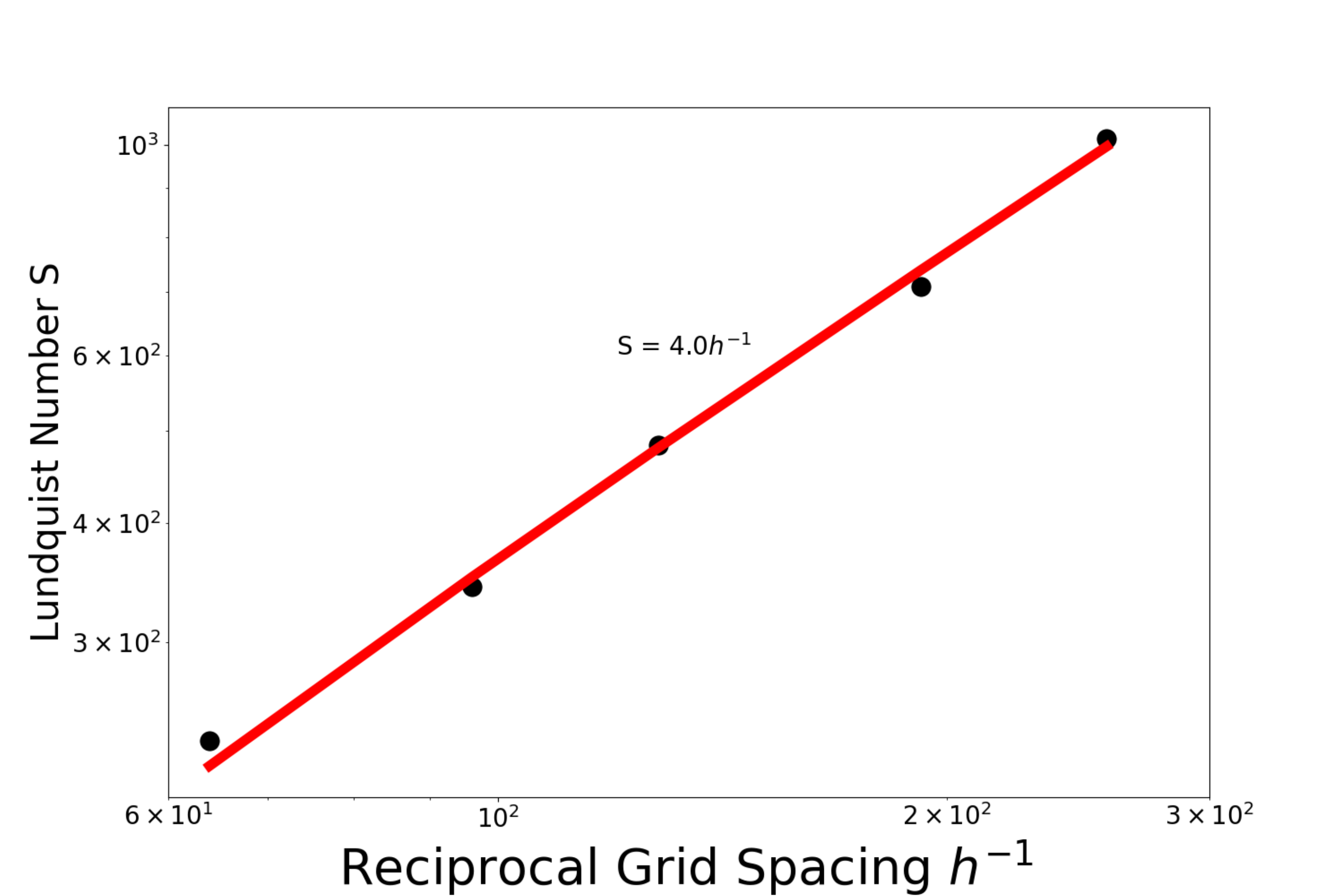}
\vspace{0mm}
\caption{Average Lundquist number versus reciprocal grid spacing ($h^{-1}$) for the three simulations with $l=2,3,4$, along with two intermediate resolution cases $l \approx 2.6, 3.6$. Also plotted is the line of best fit for the Lundquist number versus $h$.}
\label{fig:lundquist_ref}
\end{figure*}

To analyze and apply our results to the heating of the real corona, we introduce
an effective numerical diffusivity into the induction equation $\eta$,
\beg{res_induction}
\pd{\vecB}{t} =
\curl{\left(\vecv\times\vecB\right)} -
\curl{\left(\eta\curl{\vecB}\right)}.
\done
In our simulation, the energy conversion also involves dissipation of
small-scale mass motions into heat via numerical viscosity. For the
purposes of this total energy analysis, we simply lump all the dissipation
into the effective resistivity. Taking the numerical heating to be
due to an effective resistive dissipation at the reconnection sites,
the rate of change of internal energy is
\beg{res_internal_energy}
\begin{split}
\frac{dI}{dt} &= \frac{1}{4\pi} \int_V{\eta \left\vert \curl{\vecB} \right\vert^2 dV} \\ \nonumber
              &= \frac{\langle \eta \rangle}{4\pi} \int_V{\left\vert \curl{\vecB} \right\vert^2 dV}.
\end{split}
\done
We solved this equation for the spatially averaged diffusivity,
$\langle \eta \rangle$, averaged the rate of internal energy change
over the 20 driving cycles of the simulation,
\beg{res_avg_internal_energy}
\langle \frac{dI}{dt} \rangle = \frac{\Delta I^{20}_{0\%}}{20\tau},
\done
and calculated the
(quasi-periodic) maximum volume integral of
$\left\vert \curl{\vecB} \right\vert^2$ late in the simulations.
The result is
\beg{res_avg_resistivity}
\langle \eta \rangle = 4\pi
\frac{\Delta I^{20}_{0\%}}{20\tau} \left[ \langle \int_V{\left\vert
      \curl{\vecB} \right\vert^2 dV} \rangle \right]^{-1}.
\done
We then used these values to calculate the average Lundquist numbers
$\langle S \rangle$ for the simulations,
\beg{res_avg_lundquist}
\langle S \rangle = \frac{V_A L_x}{\langle \eta \rangle} =
\frac{1}{\langle \eta \rangle},
\done
in which the background Alfv\'en
speed $V_A=1$ and the current-sheet length is assumed to be the
simulation box height, $L_x=1$. The resulting data points are displayed
in Figure \ref{fig:lundquist_ref} for the Lundquist number
$\langle S \rangle$  vs.\ the reciprocal grid
spacing.  Also shown as a red line is a
linear fits to these points,
\beg{res_lundquist_fits}
\langle S \rangle \approx 4.0 h^{-1},
\done
or, alternatively, 
\beg{res_eta_fit}
\langle \eta \rangle \approx 0.25 h.
\done 
The point for the lowest refinement level, $l=1$, is omitted from
the graph and the curve fit because the grid is so coarse that the result
deviates significantly (though not anomalously) from the trends set by
the highest three levels, $l=2,3,4$. In order to fill out the interval,
we performed additional simulations at intermediate resolutions of 24
and 48 grid points per rotation cell, corresponding to refinement
levels $l \approx 2.6, 3.6$. The results obtained indicate
that the effective diffusivity scales in direct proportion to the grid
spacing, as expected for the resolution-dependent numerical diffusion
coefficients employed by our algorithm \citep{DeVore91}. Consequently,
the associated Lundquist number scales inversely with the grid
spacing. \par

\begin{figure*}[!p]
\centering\includegraphics[scale=0.5, trim=0.0cm 0.0cm 0.0cm 0.0cm]{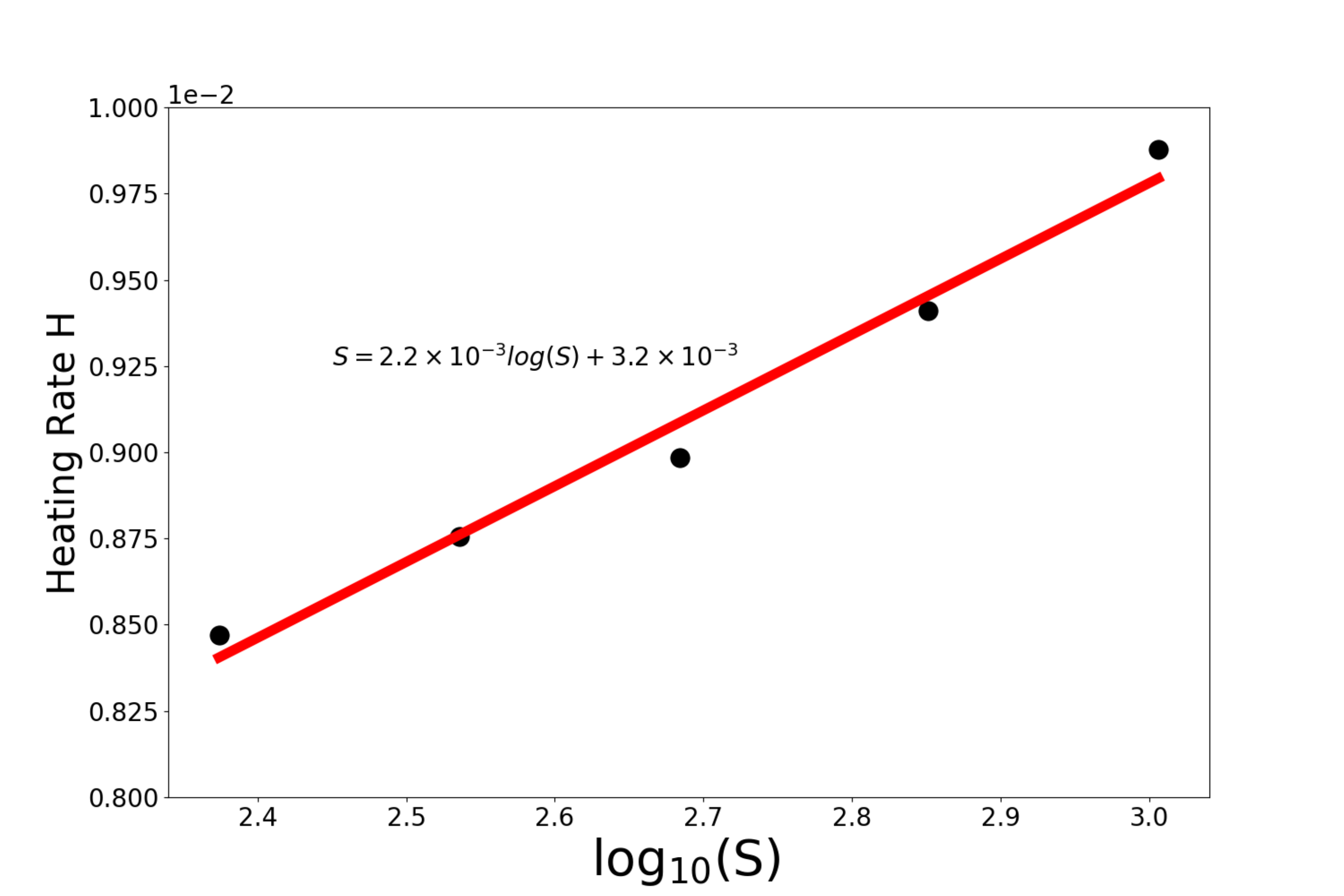}
\vspace{0mm}
\caption{Heating rate versus Lundquist number for the three simulations with $l=2,3,4$, as well as two intermediate-resolution cases $l \approx 2.6, 3.6$, along with a line of best fit.}
\label{fig:heating_ref}
\end{figure*}

For each simulation, we calculated the effective heating rate
$\langle H \rangle$ (internal energy per unit area per unit time) by
dividing the average rate of change of internal energy by the total
area (top and bottom plates) occupied by the rotation cells,
\beg{res_avg_heat}
\langle H \rangle = \frac{1}{2 N \pi a_0^2} \langle\frac{dI}{dt} \rangle.
\done
These results are displayed vs.\ the
effective Lundquist number $\langle S \rangle$ in Figure
\ref{fig:heating_ref} as filled circles. We fit a logarithmic curve to
these data, obtaining
\beg{res_avg_heat_fit}
\langle H \rangle = 2.2\times10^{-3} (\log S)+3.2\times 10^{-3},
\done
shown as the solid line in the figure. The fit here is not as tight as
those shown in the preceding figure, but it is still reasonably
good. This result allows us to extrapolate our numerical results to
coronal Lundquist numbers, as described below. \par


\section{Solar Coronal Heating}\label{sec:corona}
The results above can be related quantitatively to observed properties
of solar coronal heating. In a statistically steady state, \citet{Parker83}
relates the average coronal heating rate in a system like ours to the
inclination angle $\theta$ of the field with respect to the vertical,
\beg{Klim15}
\langle H \rangle = 
\frac{1}{4\pi} \langle v_h B_n^2 \tan \theta \rangle,
\done
where $v_h$ is the horizontal driving velocity and $B_n$ is the field component
normal to the surface. Solving for the inclination angle, known as the
Parker angle \citep{Klimchuk15}, we obtain
\beg{Parkangle}
\langle \theta_P \rangle = \arctan \left( \frac{4\pi \langle H
    \rangle}{\langle v_h B_n^2 \rangle} \right).
\done
Substituting simulation parameters
$\langle v_h \rangle \approx 0.5 v_h^{\rm max} = 0.1$,
$B_n = \sqrt{4\pi}$, and $\langle H \rangle = 0.010$ for $p=0\%$
at level $l=4$, we find
\beg{thetaP}
\langle \theta_P \rangle = \arctan \left( 0.10 \right) \approx 6^\circ.
\done
This angle is far smaller than the value
$\theta_P = \arctan \left (0.4 \right) = 22^\circ$ estimated by
\citet{Parker83} to account for the empirically measured heating of
the corona. However, we showed above that the results of our
grid-resolution study yield a logarithmic scaling law for the heating
rate, $\langle H \rangle$, in Eq.\ (\ref{res_avg_heat_fit}) and
Figure \ref{fig:heating_ref}, vs.\ Lundquist number $S$. Extrapolating
from our simulated Lundquist numbers $S \sim 10^{3-4}$ to a coronal
Lundquist number $S_\odot \sim 10^{12}$, we obtain the much larger
nondimensionalized heating rate
\beg{PoyntingExtrap}
\langle H \rangle \approx 0.030.
\done
Substituting this result into Eq.\ (\ref{thetaP}) for the
Parker angle, we find
\beg{thetaPextrap}
\langle \theta_{P \odot} \rangle = \arctan
\left( 0.30 \right) \approx 17^\circ.
\done
This is much closer to
Parker's original estimate of $22^\circ$ for the corona. \par

To convert our simulated coronal heating rate to solar values, from
Eq.\ \ref{Klim15} we have
\beg{SsimNorm}
\frac{\langle H_\odot \rangle}{\langle v_\odot B_\odot^2 \rangle} = 
\frac{\langle H \rangle}{\langle v_h B_n^2 \rangle} = 
\frac{1}{4\pi} \langle \tan \theta_{P \odot} \rangle \approx 0.024
\done
after substituting our extrapolated Parker angle $\theta_{P \odot}$
for coronal Lundquist numbers $S_\odot$ from Eq.\ (\ref{thetaPextrap}). 
The coronal heating rate therefore is
\beg{relateSsimtoSsun}
\langle H_{\odot} \rangle \approx 0.024 \langle v_\odot B_\odot^2 \rangle.
\done
Assuming a photospheric flow speed $v_{\odot} = 1 \times 10^5$ cm
s$^{-1}$ and an average quiet-Sun flux density $B_{\odot} = 10$ G, we
obtain
\beg{SsunQS}
\langle H_{\odot} \rangle_{QS} \approx 2.4 \times 10^5 \;
  \mathrm{erg\; cm^{-2}\; s^{-1}}.
\done
This value is in very good agreement with accepted quiet-Sun energy
loss rates \citep{Withbroe77}. In active regions, on the other hand,
typical average magnetic flux
densities are an order of magnitude larger than those in quiet Sun
\citep[e.g.][]{MP97}, so the Poynting flux density is two orders of
magnitude larger than the quiet-Sun value above. The resulting heating
rate,
\beg{SsunAR}
\langle H_\odot \rangle_{AR} \approx 2.4 \times 10^7 \;
  \mathrm{erg\; cm^2\; s^{-1}},
\done
also is in excellent agreement with the observed
rate of energy loss from active regions \citep{Withbroe77}. \par


\section{Discussion}\label{sec:discussion}
In this work, we calculated the effects of helicity injection on the
heating of the corona using helicity- and energy-conserving numerical
simulations. Helicity was injected into the corona via numerous
convective cells at the photosphere, whose sense of rotation and
position were randomly varied. In previous papers, we showed
that this helicity was transported throughout the simulation domain by
magnetic reconnection, forming filament channels (KAD15) and leaving
behind a relatively smooth corona (KAD17). In this paper, we explored
how magnetic reconnection distributes the injected energy between the
magnetic field and the plasma, and we found a number of important
results. \par

The main conclusion is that, irrespective of helicity, the majority of
the injected energy is converted into heating, with the rest building up
the free energy of the coronal magnetic field. Furthermore, we found that
the amount of heating depends only weakly upon the helicity preference. 
It seems unlikely, therefore, that the helicity preference of the
photospheric motions can be inferred from measurement of the coronal
heating rate.  On the other hand, the amount of energy remaining in
the magnetic field depends strongly upon this preference, with the
most magnetic energy remaining in the field for larger helicity
preferences. Given the observed ubiquitous formation of filament
channels in the corona, along with the frequent coronal mass
ejections, it seems highly likely that the photospheric convective
motions inject a net helicity into the corona \citet[e.g.][]{Mackay18}. 
Furthermore, our results imply that current-sheet formation and
reconnection are most efficient for maximal helicity preferences and
least efficient for negligible helicity preferences. This conclusion is
in line with previous studies of flux tube dynamics \citep[e.g.][]{Zhao15}. 
Crucially, we also showed quantitatively that the resulting energy flux
into the corona can account for the observed
heating. This result agrees with the many studies, dating back to
\citet{Parker72}, that the observed photospheric convective flows and
photospheric field strengths are sufficient to heat the corona. \par

Taken together with KAD15 and KAD17, the results of this paper
constitute a comprehensive model for helicity injection into the
corona. In this model, surface motions inject free energy and helicity
into the solar corona. The helicity is transported to the largest
available scales by magnetic reconnection where it is lost only
through occasional explosive ejections, while most of the energy is converted
to heat. The single process of magnetic helicity condensation, therefore,
can simultaneously explain the formation of filament channels that
lead to coronal ejections, the smoothness of coronal loops, and the
heating of the Sun's multi-million degree corona. \par

Our study also suggests that the reconnection responsible for helicity
condensation should be explored in more detail. Does such reconnection
proceed as a series or, perhaps, a storm of nanoflares \citep{Klimchuk15}?
Or is the reconnection, instead, a single, spontaneous event? Another
important issue to be explored is the relation between the size of the
individual reconnection events -- the nanoflares -- and the properties 
of the photospheric driving. 
{\color{black} Future work also should address the influence of the 
magnetic topology, in configurations having true PILs such as the 
bipolar sunspot simulated by \citet{Knizhnik17b}, on the rate and 
distribution of coronal heating.}
Irrespective of the resolution of these issues,
however, it is clear from our study, and many others, that the injection
and transport of magnetic energy and helicity play major roles in determining
the structure, dynamics, and heating of the solar corona. \par


\acknowledgments{K.J.K.\ gratefully acknowledges funding for this
  work received through a NASA Earth and Space Science Fellowship, as
  well as the use of codes written by B.\ J.\ Lynch. The numerical
  simulations were performed under a grant of NASA High-End Computing
  resources to C.R.D.\ on Discover at NASA's Center for Climate
  Simulation. Grants
  from NASA's Living With a Star, Heliophysics Supporting Research, 
  and Heliophysics Internal Scientist Funding Model programs partially 
  supported K.J.K., S.K.A., J.A.K., and C.R.D.}


\newpage

\newpage

\newpage

\end{document}